\documentclass[onecolumn,journal,draftclsnofoot,12pt]{IEEEtran}

\usepackage{bm}
\usepackage{amssymb}
\usepackage{amsmath}
\usepackage{amsfonts}
\usepackage{graphicx}
\usepackage{subfigure}
\usepackage{cite}
\usepackage{enumerate}
\usepackage{url}
\usepackage{epstopdf}
\usepackage{float}
\usepackage{color}
\usepackage{mathtools}
\usepackage{multirow}
\usepackage{enumitem}
\usepackage{enumerate}
\usepackage[rgb]{xcolor}
\usepackage{enumitem}

\usepackage[normalem]{ulem}  

\newcommand{\beq}{\begin{equation}}
\newcommand{\eeq}{\end{equation}}

\allowdisplaybreaks[4]

\makeatletter  
\newif\if@restonecol  
\makeatother

\usepackage{amsmath}

\usepackage{algpseudocode}
\usepackage{algorithmicx,algorithm}

\usepackage{makecell} 

\makeatletter
\newcommand{\manuallabel}[2]{\def\@currentlabel{#2}\label{#1}}
\makeatother

\DeclareMathOperator*{\argmax}{argmax}

\newif\ifshowannote
\showannotetrue

\newtheorem{proposition}{\bf Proposition}

\usepackage[numbers,sort&compress]{natbib}
\usepackage{etoolbox}
\makeatletter
\patchcmd{\NAT@citexnum}
  {%
    \ifx\NAT@last@yr\relax
      \def@NAT@last@yr{\@citea}%
    \else
      \def@NAT@last@yr{--\NAT@penalty}%
    \fi
  }
  {%
    \def@NAT@last@yr{--\NAT@penalty}%
  }
  {}{\FAIL}
\newcommand{\rcsf}{reflection coefficient function}

\newcommand{\mms}{meta-IoT sensor}
\newcommand{\Mms}{Meta-IoT sensor}
\newcommand{\MmS}{Meta-IoT Sensor}

\newcommand{\mm}{meta-material}

\newcommand{\mmsd}{meta-IoT structure}
\newcommand{\Mmsd}{Meta-IoT structure}
\newcommand{\MmsD}{Meta-IoT Structure}

\newcommand{\mmsu}{{\mms} unit}
\newcommand{\mmu}{meta-IoT unit}

\newcommand{\senscond}{environmental condition}
\newcommand{\envircondvec}{environmental condition vector}

\newcommand{\rfmeasure}{wireless transceiver}
\newcommand{\RfMeasure}{Wireless Transceiver}

\newcommand{\corenetname}{encoder-decoder neural network}

\newcommand{\CoreNetName}{Encoder-decoder Neural Network}

\renewcommand{\Pr}{\mathrm{Pr}}
\renewcommand{\d}{\mathrm{d}}

\newcommand{\erf}{\mathrm{erf}}
\newcommand{\iu}{\mathit{i}}
\newcommand{\cen}{\mathrm{c}}
\newcommand{\m}{\mathrm{M}}
\newcommand{\s}{\mathrm{s}}
\newcommand{\w}{\mathrm{w}}
\newcommand{\T}{\mathrm{T}}
\newcommand{\F}{\mathrm{F}}
\newcommand{\tx}{\mathrm{Tx}}
\newcommand{\rx}{\mathrm{Rx}}
\newcommand{\MS}{\mathrm{MS}}

\newcommand{\AT}{\mathrm{AT}}

\begin{document}

\title{{Meta-material Sensor Based Internet of Things: Design, Optimization, and Implementation}}

\author{
\IEEEauthorblockN{
\normalsize{Jingzhi~Hu},~\IEEEmembership{\normalsize Graduate~Student~Member,~IEEE},
\normalsize{Hongliang~Zhang},~\IEEEmembership{\normalsize Member,~IEEE},
\normalsize{Boya~Di},~\IEEEmembership{\normalsize Member,~IEEE},
\normalsize{Zhu~Han},~\IEEEmembership{\normalsize Fellow,~IEEE},
\normalsize{H.~Vincent~Poor},~\IEEEmembership{\normalsize Life~Fellow,~IEEE},
and~\normalsize{Lingyang~Song},~\IEEEmembership{\normalsize Fellow,~IEEE}
}
\thanks{
 J. Hu, B. Di, and L. Song are with School of Electronics, Peking University, Beijing 100871, China~(email: \{jingzhi.hu, diboya, lingyang.song\}@pku.edu.cn).
 }
 \thanks{
H. Zhang and H. V. Poor are with Department of Electrical Engineering, Princeton University, Princeton, NJ 08540 USA~(email: hongliang.zhang92@gmail.com, poor@princeton.edu).
}
\thanks{
Z. Han is with Electrical and Computer Engineering Department, University of Houston, Houston, TX 77004 USA, and also with Department of Computer Science and Engineering, Kyung Hee University, Seoul 446-701, South Korea~(email: hanzhu22@gmail.com).
}
}

\maketitle
\vspace{-2em}
\begin{abstract}
For many applications envisioned for the Internet of Things (IoT), it is expected that the sensors will have very low costs and zero power, which can be satisfied by meta-material sensor based IoT, i.e., meta-IoT.
As their constituent meta-materials can reflect wireless signals with environment-sensitive reflection coefficients, meta-IoT sensors can achieve simultaneous sensing and transmission without any active modulation.
However, to maximize the sensing accuracy, the structures of meta-IoT sensors need to be optimized considering their joint influence on sensing and transmission, which is challenging due to the high computational complexity in evaluating the influence, especially given a large number of sensors.
In this paper, we propose a joint sensing and transmission design method for meta-IoT systems with a large number of meta-IoT sensors, which can efficiently optimize the sensing accuracy of the system.
Specifically, a computationally efficient received signal model is established to evaluate the joint influence of meta-material structure on sensing and transmission.
Then, a sensing algorithm based on deep unsupervised learning is designed to obtain accurate sensing results in a robust manner.
Experiments with a prototype verify that the system has a higher sensitivity and a longer transmission range compared to existing designs, and can sense environmental anomalies correctly within $2$ meters.
 
\end{abstract}


\section{Introduction}

\IEEEPARstart{U}{nderstanding} the surrounding environments of humans and their development is of crucial importance for improving the quality of life and promoting work efficiency~\cite{Othman2012Wireless}.
For this purpose, in the upcoming 6G communications, it is envisioned that an extremely large number of Internet of Things~(IoT) sensors will spread to collect information of living environments, which will be approximately 10-fold more than that in 5G~(10 million devices per square km)~\cite{Zhang2020Beyond}.
Among them, a massive number of IoT sensors will be embedded for sensing applications to monitor environmental conditions, structural integrity, and pervasive industrial processes~\cite{Wikstrom2020Challenges}.
In such sensing applications, it is required that the IoT systems with large numbers of sensors should be cost and energy efficient, easy to maintain, and environmentally sustainable~\cite{Huang2019ASurvey}.

To satisfy the above requirements, the IoT in 6G is expected to have sensors with extremely low cost, zero-power consumption, and no human intervention demands (e.g. recharging and maintenance)~\cite{Wikstrom2020Challenges}.
Researchers have been making constant efforts in designing low-power~(LP) IoT sensors following the standards such as Zigbee, Bluetooth and its low energy extension, LoRa, and NB-IoT~\cite{Zanaj2021Energy}.
However, these LP IoT sensors still need power supplies like batteries to support their sensing and transmission.
Though each LP IoT sensor only consumes mW level power in their active phase~\cite{Lauridsen2018AnEmpirical}, their collective power consumption can be considerable in the massive deployment.
Besides, as the battery capacity is limited, typically LP IoT sensors have a lifetime of around three to five years~\cite{Nikoukar2018Low}{}, and the periodic battery changing task leads to high maintenance costs.

Recently, various radio-frequency identification (RFID) sensors have been designed for sensing applications, which can transmit sensory data by reflecting RF signals without additional supplied power~\cite{Costa2021AReview,Mezzanotte2021Innovative}.
Nevertheless, they still require sophisticated RF energy harvesters and microchips for sensing and signal modulation, which makes their cost unable to be cheap enough for massive deployment~\cite{Costa2021AReview}.
Moreover, the microchips in the RFID sensors are likely to encounter unexpected electrical failures and breakdowns when their service time is long, especially when they are deployed in harsh environments~\cite{Ozturk2019RFID}{}.
In this case, human intervention and maintenance are also inevitable which will incur heavy tasks of finding and replacing failed sensors.
As a result, current LP IoT sensors and RFID sensors are not suitable for the massive deployment in 6G IoT.

In contrast, \emph{\mms}s, i.e., sensors composed of {\mm}s, are attractive candidates for IoT sensors in 6G.
The {\mms}s achieve simultaneous sensing and transmission by signal reflection on metamaterial structures and are free of expensive and delicate modulation microchips and power supplies, making them significantly more cost/energy-efficient and robust.
Specifically, {\mms}s are composed of sub-wavelength split-ring resonator~(SRR) structures that are referred to as \emph{\mmsd}s and have high sensitivity towards environmental conditions.
When wireless signals of a certain frequency impinge on the {\mmsd}, the signals are reflected and become dependent on surrounding environmental conditions.
Therefore, by reflecting wireless signals, the {\mms}s enable environment sensing and signal transmission towards a distant receiver at the same time, and thus they follow a joint sensing and transmission working scheme.

Typically, meta-IoT sensing systems are comprised of {\mms}s and wireless transceivers.
In this configuration, the sensing results are obtained by the receiver using a dedicated sensing algorithm, which should consider the joint influence of sensing and transmission on the received signals.
The main difference between the meta-IoT system and traditional IoT sensing systems is that it adopts metamaterial structure based passive chipless sensors.
This results in that the sensing results are obtained by the receiver with received signal processing as in RF sensing systems like~\cite{Hu2020Reconfigurable, Zhang2022Towards} rather than by the sensors themselves.
 
However, it remains a challenge to design an efficient meta-IoT system.
\emph{Firstly}, designing the {\mmsd} for improving sensing accuracy is challenging.
This is because it requires a computationally efficient model to capture the joint influence of the environmental conditions and the {\mmsd} on the received signals.
Such a model is difficult to establish especially in situations with larger numbers of {\mms}s as expected in 6G.
\emph{Secondly}, the properties of wireless transmission, such as the multi-path effect and thermal noises, need to be considered particularly in designing the sensing algorithm.
This is because, in the received signals, the influence of wireless transmission interferes with that of the environmentally sensitive reflection on the {\mms}s, which makes it non-trivial for the receiver to infer the environmental conditions accurately.
However, existing works lack efficient methods to optimize the structure of {\mm}s and have not considered the impairments of wireless transmission in designing the sensing algorithms, which result in a limited transmission range.

To address the above challenges, we develop a joint sensing and transmission design for general meta-IoT systems with large numbers of meta-IoT sensors.
We focus on the design of the meta-IoT structure and received signal processing problem since they are the main features of meta-IoT systems.
Specifically, we propose an analytical model of the received signals, which jointly considers the influence of the meta-IoT structure on the sensing and transmission for scenarios with large numbers of meta-IoT sensors. Based on this signal model, we optimize the meta-IoT structure for a general objective which is suitable for various sensing applications. Then, aiming at anomaly detection applications, we design a sensing algorithm for the system to obtain sensing results from received signal processing. The designed sensing algorithm is based on unsupervised learning and an encoder-decoder neural network, which can handle the properties of transmission such as the multi-path effect and thermal noises.
Furthermore, a prototype of the designed system is implemented and tested in a typical indoor environment. 
In summary, the main contributions of this paper are:
\begin{itemize}[leftmargin=*]
\item 
We propose a received signal model for {\mmsd} optimization in the meta-IoT systems with a large number of {\mms}s. 
Compared with existing works using full-wave simulation, the proposed analytical model largely reduces the computational complexity in evaluating the influence of {\mmsd} on joint sensing and transmission.
\item 
We design a deep unsupervised learning based sensing algorithm for meta-IoT systems. In contrast to the existing works based on direct comparison, the proposed algorithm can effectively handle the impairments due to wireless transmission and obtain accurate sensing results robustly.
\item
We implement a prototype of the designed system, which achieves higher sensitivity and longer transmission range compared with the existing works in a similar frequency band and having similar functions.
Specifically, in a typical indoor setting, the prototype system can detect and locate temperature and humidity anomalies successfully with a range of $2$ m.
\end{itemize}

The rest of the paper is organized as follows.
In Section~\ref{sec: relate work}, we discuss related work, and in Section~\ref{sec: meta sensor module}, we provide an overview of {\mms}s.
Then, in Section~\ref{sec: overview}, we describe our system model and propose a received signal model.
Following that, we formulate and solve the joint sensing and transmission optimization problem for {\mms}s in Section~\ref{sec: joint sensing and transmission optimization}.
In Section~\ref{sec: anomal detect and locate mod}, we propose a sensing algorithm that is suitable for the meta-IoT systems to sense environmental anomalies.
In Section~\ref{sec: system implementation}, we present our prototype of the designed system and explain its implementation.
In Section~\ref{sec: evaluation}, experimental results for the proposed system are provided.
Finally, Section~\ref{sec: conclu} draws our conclusions.
 
\section{Related Work}
\label{sec: relate work}

In this paper, we focus on meta-IoT sensing systems and their design methods.
The main difference between the meta-IoT sensing systems and existing IoT systems is that meta-IoT sensing systems employ meta-IoT sensors which can perform joint sensing and transmission.

Therefore, in this section, we first review the existing works on the sensors that perform joint sensing and transmission, which are the \emph{chipless passive sensors}.
By comparing them with the meta-IoT sensors, we point out the advantages of the meta-IoT sensors and meta-IoT sensing systems.
Secondly, we summarize the existing meta-material sensor based systems and discuss the inadequacies in their designs.

\subsection{Chipless Passive Sensors}
Recently, some chipless passive wireless sensors have been proposed to sense environmental conditions, which can perform joint sensing and transmission.
In~\cite{Vena2014AFully}, the authors proposed a passive wireless sensor composed of SRR circuits and a climate-sensitive polymer, which can be used to sense CO$_2$ concentration or temperature within a gas container.
In~\cite{Hester2016Inkjet}, the authors designed a conductive-ink-based passive humidity sensor printed on a flexible Kapton film and used a focused millimeter-wave wireless transceiver to enable a long transmission range.
In~\cite{Amin2013ANovel}, the authors designed a chipless RFID tag to sense humidity, which is composed of a patch with slot resonators and an electric inductive-capacitive resonator.
In~\cite{Deng2018Design}, the authors also proposed a chipless RFID sensor tag for humidity monitoring, which is composed of slotted scatterer structures fabricated on the substrate of a printed circuit board.

However, traditional chipless passive sensors have larger sizes compared with their working wavelengths.
Then, when multiple traditional passive sensors form a sensor array to support ubiquitous sensing applications in 6G, their unit spacing interval is larger than the working wavelength.
Consider the passive sensors as antennas, and then, based on the antenna array theory~\cite{goldsmith2005wireless}, the array of traditional passive sensors with a larger unit spacing interval than the working wavelength has a poor concentration of reflected energy, which shortens the transmission rage.
As a result, without the help of expensive mmWave equipment to focus wireless signals~\cite{Vena2014AFully} and increase the incident power, the traditional chipless passive sensors have a short transmission range.

Compared with traditional chipless passive sensors, the {\mms}s are composed of densely paved {\mmsu}s with a smaller size than half of the working wavelength, which provides {\mms}s with two features.
Firstly, it enables the {\mms}s to have a larger transmission range, since their reflected signal power is more concentrated according to the antenna array theory~\cite{goldsmith2005wireless}.
Secondly, for a given area, a larger number of {\mmsu}s can be filled into the area, which indicates a higher sensitivity.

\subsection{Meta-meterial Sensor Based  Systems}
Several existing {\mm} sensors have been proposed to measure physical conditions such as temperature and humidity.
In~\cite{Karim2017Metamaterial}, the authors proposed to sense temperature with an array of {\mm}-based wireless sensors, and demonstrated the feasibility of wireless temperature sensing up to $200^\circ$C.
In~\cite{Ekmekci2019TheUse}, the authors designed a {\mm}-based sensor to sense humidity and methanol-deionized water solutions inside a waveguide testbed.
In~\cite{Hu2021Meta-IoTGC, Hu2021Meta-IoTTWC}, the authors designed a general {\mm}-based sensor for environmental conditions, such as temperature and humidity, and proposed an algorithm to optimize the structure of the sensors to enhance the sensing accuracy.
Moreover, in~\cite{Liu2021Meta-IoTGC}, the authors used multiple {\mm} sensors and a supervised learning algorithm to sense the distribution of environmental conditions.

Nevertheless, existing {\mm}-based wireless sensing systems only consider a small number of sensors, whereas 6G demands a large or even massive number of sensors to be deployed.
The methods designed for the cases with a small number of sensors cannot be extended to the cases with a large number of sensors due to the following reasons.
In the small number of sensors case, the received signals can be modeled by using full-wave simulation or the summation of the reflected signals from each sensor. 
Besides, as the deployment scale is small, the environment around sensors can be controlled. Thus, a data set of received signals labeled with known environmental conditions can be obtained, which enables empirical comparison and supervised learning-based sensing algorithms.
However, when the number of sensors is large, the received signal model in the small number of sensors case leads to high computational cost.
Moreover, controlling the environment around sensors becomes infeasible, and thus an unsupervised sensing algorithm is in need.

Furthermore, existing works have not considered the impairment of wireless transmission in their system design, such as the effects of multi-paths and noises.
Since both the impairments of wireless transmission and the environmentally sensitive reflection on the {\mms}s impact the received signals, their mutual interference can make the receiver hard to infer the environmental conditions correctly.
To handle the above issues, an efficient sensing algorithm needs to be designed.

\section{{\Mms}s}
\label{sec: meta sensor module}

In this section, we introduce the meta-IoT sensors, which are the core component of meta-IoT systems.
We first elaborate on the structure of meta-IoT sensors, and then establish a reflection coefficient model of meta-IoT sensors.
By using the equivalent circuit model, the reflection coefficients of meta-IoT sensors are shown to be sensitive to environmental conditions and can be designed by changing metamaterial structures, which provides insight into meta-IoT system design.

\subsection{{\MmsD}}
\label{ssec: mmr design}

{\Mms}s are based on {\mm}s, which are artificial periodic structures with exotic properties that cannot be found in nature~\cite{Munk2000Frequency}.
These exotic physical properties exhibited by {\mm}s are underpinned by their special frequency responses.
Specifically, when wireless signals impinge upon a {\mms}, its frequency response is determined by the {\mmsd} and the conditions of surrounding environments.

\begin{figure}[!t]
\center{\includegraphics[width=0.5\linewidth]{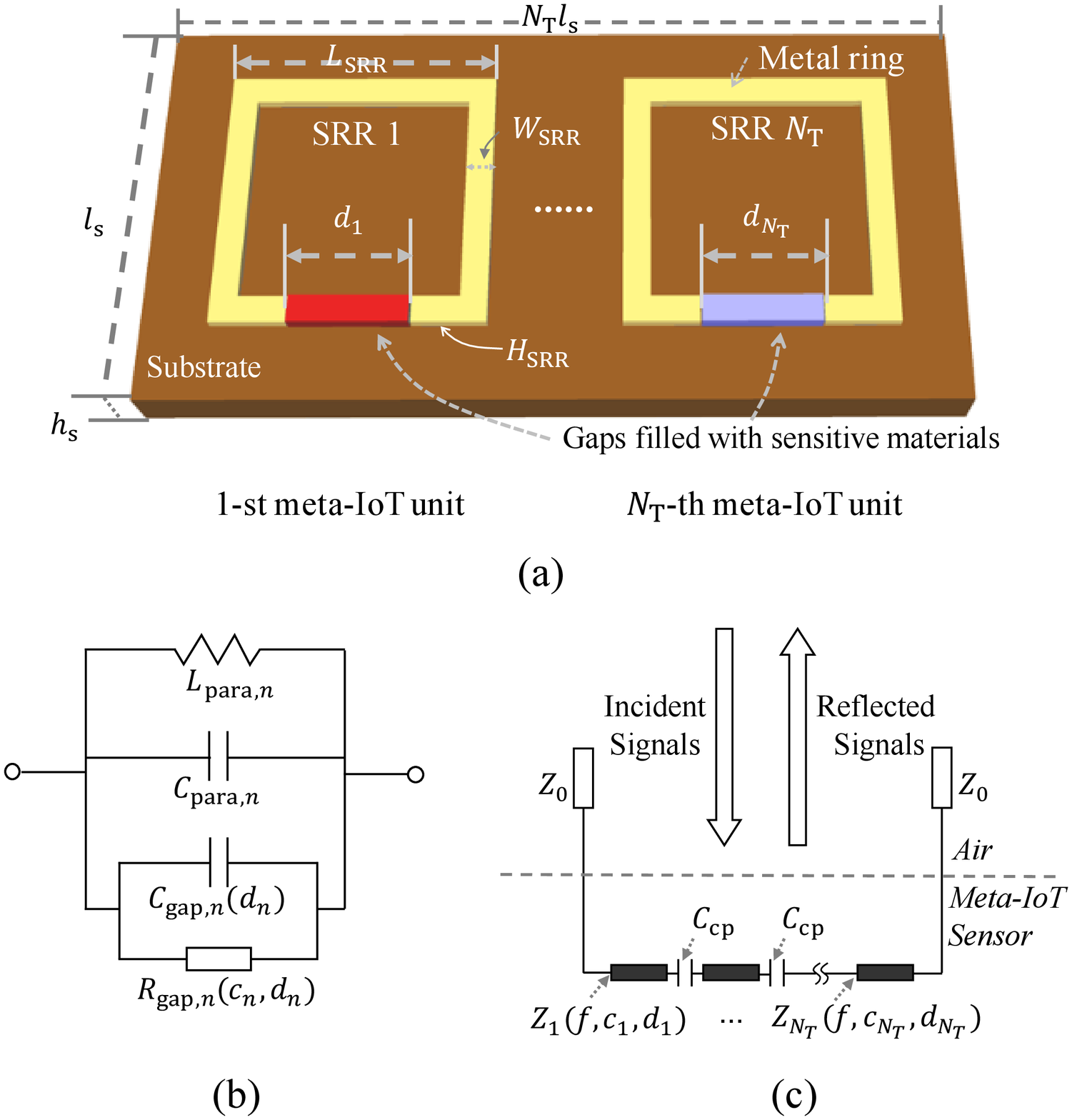}}
	\vspace{-.5em}
	\setlength{\belowcaptionskip}{-1.em} 	\caption{(a) A {\mms} for sensing $N_\mathrm{T}$ {\senscond}s. (b) Equivalent circuit model of the $n$-th {\mmsu}. (c) Equivalent circuit model of the {\mms}.}
	\label{fig: sensor overview}
\end{figure}

A {\mms} consists of $N_\mathrm{T}$ units for sensing $N_\mathrm{T}$ different environmental conditions.
Each {\mmu} consists of a metal SRR with a horizontal gap printed on a square supportive dielectric substrate with side-length $l_s$ and thickness $h_s$.
Specifically, as illustrated in Fig.~\ref{fig: sensor overview}~(a), the metal ring has the side length $L_{\mathrm{SRR}}$, and its width and thickness are denoted by  $W_{\mathrm{SRR}}$ and $H_{\mathrm{SRR}}$, respectively.
The gap of the SRR is of width $d_n$ and is filled with sensitive materials, whose electric properties are sensitive to the $n$-th environmental condition.
Due to the sensitive materials, a {\mms}'s frequency response, i.e., reflection coefficients for wireless signals, are sensitive to the $N_\mathrm{T}$ {\senscond}s.

\subsection{Reflection Coefficient Model}
\label{ssec: ref coeff model}
In the following, we model the reflection coefficient of a {\mms}, which is necessary for establishing the model of received signals.
Firstly, based on~\cite{Vena2014AFully}, an SRR is only excited by the RF signals having the same polarization direction as its gap direction, which means the signals with cross-polarization cannot lead to environmentally sensitive resonant absorption of SRRs.
Thus, in this paper, we assume Tx and Rx antennas in the meta-IoT system to be linearly polarized and align their polarization directions with the meta-IoT sensors' gap direction.
In this case, most of the incident signals on {\mms}s have the electric field intensity parallel to the gap, and the signals with cross-polarization have little influence on the received signals.

Based on the transmission line theory in~\cite{pozar2011microwave}{}, we model the meta-IoT sensor as an equivalent load impedance connected to an equivalent transmission line of the free space, which is a commonly adopted model for meta-material particles in communication scenarios~\cite{Tang2019Wireless2}.
As the SRR structure of each meta-IoT unit is a resonant circuit for incident RF signals, we model the equivalent load impedance of each meta-IoT unit as an RLC resonant circuit shown in Fig.~\ref{fig: sensor overview}~(b), where R, L, and C indicate the resistance, inductance, and capacitance, respectively~\cite{withayachumnankul2013metamaterial,Kairm2015Concept}.
For the $n$-th {\mmu} with gap width $d_{n}$, given the $n$-th environmental condition being $c_n$, the impedance of the circuit can be calculated as
\begin{align}
\label{equ: impendance of v-SRR}
Z_n(f, c_{n}, d_{n})  \!=\big( 
{1\over 2\pi\iu f L_{\mathrm{para},n}} + 
{2\pi\iu f C_{\mathrm{para},n}}  +{2\pi\iu f C_{\mathrm{gap},n}(d_n)} 
+ {1\over R_{\mathrm{gap},n}(c_n, d_n)}
\big)^{-1}, 
\end{align}
where $\iu$ denotes the imaginary unit,
$f$ denotes the frequency of the signals, and
$L_{\mathrm{para},n}$ and $C_{\mathrm{para},n}$ are the parasitic inductance and capacitance of the SRR, respectively.
Besides, based on~\cite{Jones2013The}, the resistance $R_{\mathrm{gap},n}(c_n, d_n)$ and capacity $C_{\mathrm{gap},n}(d_n)$ can be modeled as
\begin{subequations}
\begin{align}
\label{equ: resistance with d}
& R_{\mathrm{gap}}(c_n, d_n)  =  \frac{d_n}{\rho_{\mathrm{mat}, n}(c_n)W_{\mathrm{SRR}}H_{\mathrm{SRR}} }, \\
& C_{\mathrm{gap}, n}(d_n) = \hat{C}_{\mathrm{gap},n}/{d_n}, 
\end{align}
\end{subequations}
where $\rho_{\mathrm{mat}, n}(c_n)$ denotes the conductivity of the $n$-th sensitive material,
$W_{\mathrm{SRR}}$ and $H_{\mathrm{SRR}}$ denote the width and thickness of each SRR,
and $\hat{C}_{\mathrm{gap},n}$ denotes capacity of the gap with a unit width.

As shown in Fig.~\ref{fig: sensor overview}~(c), the total impedance of the {\mms} can be expressed as 
\beq
\label{equ: total impedance}
Z(f, \bm d, \bm c) = 
{N_T-1\over 2\pi\iu f\cdot C_{\mathrm{cp}}}
+ \sum_{n=1}^{N_T} Z_n(f, c_n, d_n),
\eeq
where $C_{\mathrm{cp}}$ denotes the capacity due to the coupling between adjacent {\mmsu}s.
Besides, $\bm c=(c_1,...c_{N_T})$ denotes the environmental condition vector. 

For the {\mms}, its \emph{reflection coefficient} is a parameter that describes the fraction of the wireless signals reflected by an impedance discontinuity in the transmission medium~\cite{pozar2011microwave}.
It can be calculated by the ratio between the electric field intensities of the incident and reflected signals.
Then, based on~\cite{pozar2011microwave}, the reflection coefficient can be analytically modeled by 
\begin{equation}
\label{equ: reflection coefficient}
\hat{\gamma}(f, \bm d, \bm c) = {Z(f, \bm d, \bm c) - Z_0\over Z(f, \bm d, \bm c) + Z_0},
\end{equation}
where $Z_0=377~\Omega$ is the characteristic impedance of the free-space transmission line.

In (\ref{equ: resistance with d}), it can be observed that the environmental conditions determine the resistance of meta-IoT units, which influences the impedance and the reflection coefficients of the meta-IoT sensor for RF signals as shown by Eqs.~(\ref{equ: impendance of v-SRR}),~(\ref{equ: total impedance}), and~(\ref{equ: reflection coefficient}).
Thus, given the received RF signals which contain the reflected signals from meta-IoT sensors, a receiver can obtain the reflection coefficients of meta-IoT sensors, based on which it can recognize the environmental conditions.

Moreover, by substituting~(\ref{equ: impendance of v-SRR})$\sim$(\ref{equ: total impedance}) into~(\ref{equ: reflection coefficient}), we can observe that the reflection coefficient of the {\mms} is dependent on $\bm d$.
Therefore, the gap widths of the $N_\mathrm{T}$ {\mmu}s can be considered as the variables to design the {\mms}, which are thus referred to as the \emph{{\mmsd} vector}.
Using the analytical model derived above, i.e., $\hat{\gamma}(f, \bm d, \bm c)$, we reveal the influence of $\bm d$ on the reflection coefficients.
Nevertheless, to obtain a precise {\rcsf}, numerical full-wave simulation and practical experiments are in need, which is of high computational time.
Therefore, to reduce the time consumption required in optimizing $\bm d$ while ensuring the effectiveness of the results, we use $\hat{\gamma}(f,  \bm d, \bm c)$ and an additional interpolation function together to fit the precise {\rcsf} over a sampled set of $\bm d$ denoted by $\hat{\mathcal D}_\mathrm{A}$ as in~\cite{Hu2021Meta-IoTTWC}.
Specifically, in this paper, we focus on the performance of meta-IoT sensors under the $N_C$ preselected environmental conditions, which are denoted by set $\mathcal C$. 
Then, for each $\bm c\in \mathcal C$, the interpolation is performed in terms of $\bm d$, and the resulting function is denoted by ${\gamma}(f,\bm d, \bm c)$, which is used in the following sections.

\section{System Model}
\label{sec: overview}

\begin{figure}[!t]
\center{\includegraphics[width=0.7\linewidth]{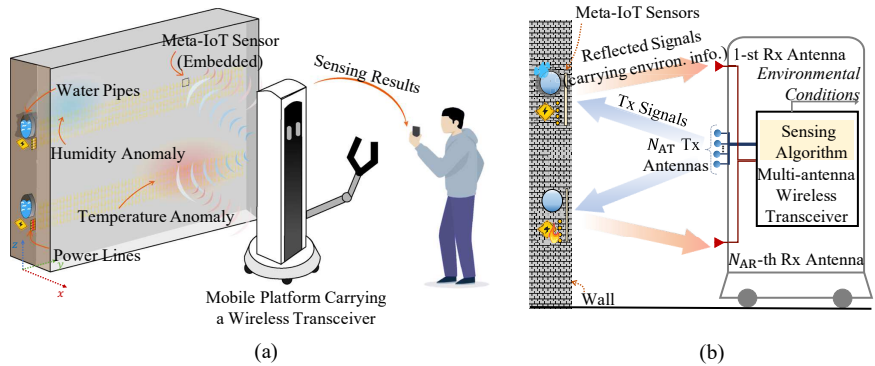}}
	\vspace{-.5em}
	\setlength{\belowcaptionskip}{-1.em}
	\setlength{\abovecaptionskip}{-.5em}
	\caption{(a) Illustration and (b) detailed schematic of an example of the meta-IoT system with $N_{\mathrm{AR}}=2$ and $N_{\mathrm{AT}}=4$. 
	The system is used to sense temperature and humidity in the target areas where power lines and water pipes are embedded.}
	\vspace{0.em}
	\label{fig: system overview}
\end{figure}

In this section, we describe the designed meta-IoT system, which is based on the meta-IoT sensors introduced in Section~\ref{sec: meta sensor module}. 
We first describe the system components and then propose the received signal model, where the reflection coefficients of meta-IoT sensors are based on the model established in Section~\ref{sec: meta sensor module}.
Specifically, computationally efficient models of received signals are proposed considering the cases where the numbers of meta-IoT sensors are small and large, respectively.

 \subsection{System Components}
 \label{ssec: sys comp}
The meta-IoT system is composed of two components, i.e., multiple {\mms}s and a multi-antenna {\rfmeasure}, which can be described as follows.

\subsubsection{{\MmS}s}
\label{s3ec: mmss}
The system contains $N_{\mathrm{AR}}$ arrays of the {\mms}s. 
The width of each array is denoted by $L_{\MS}$, and the center height of the $i$-th array is denoted by $h_i^{\MS}$.
Each {\mms}'s reflection coefficients for wireless signals are sensitive to $N_\mathrm{T}$ different environmental conditions. 
The {\mms}s are densely deployed along the wall and covering the areas that need to be sensed and monitored.
For generality, we assume that the {\mms}s are deployed in the wall at depth $D_{\mathrm{w}}$.
This assumption is fit for both embedded scenarios where $D_{\mathrm{w}}>0$ and the non-embedded scenarios where $D_{\mathrm{w}}=0$.

\subsubsection{Multi-antenna {\RfMeasure}} The {\rfmeasure} contains a Tx antenna array with $N_{\mathrm{AT}}$ antennas, $N_{\mathrm{AR}}$ Rx antenna, and a signal processor.
	The center height of the Tx antenna array is denoted by $h_{\cen}^{\mathrm{AT}}$, and the interval spacing between Tx antennas is $\delta^{\mathrm{AT}}$.
	All the Tx and Rx antennas are directional with main lobes horizontally orientated towards the wall.
	Through the Tx antennas, signals within frequency band $[f_{\mathrm{lb}}, f_{\mathrm{ub}}]$ are transmitted towards the $N_{\mathrm{AR}}$ {\mms} arrays.
	Then, the $N_{\mathrm{AR}}$ Rx antennas measure the received signals, based on which the signal processor uses a \emph{sensing algorithm} to estimate the environmental conditions.
	Moreover, the {\rfmeasure} is installed on a mobile platform, such as a home-service robot or a UAV~\cite{Zhang2019Cellular2}, which can move to sense the environmental conditions at different locations.
An example of a meta-IoT system with $N_{\mathrm{AR}}=2$ and $N_{\mathrm{AT}}=4$ is depicted in Figs.~\ref{fig: system overview}~(a) and~(b), which is used for sensing temperature and humidity inside buildings.
As shown in Fig.~\ref{fig: system overview}~(a), the {\mms}s are embedded inside the building and covering some target sensing areas (e.g., the areas where power lines and water pipes are embedded).
At a certain location shown in Fig.~\ref{fig: system overview}~(b), the wireless transceiver transmits signals that are reflected back by the {\mms}s and carrying the information of the environmental conditions.
By handling the received reflected signals with its sensing algorithm, the wireless transceiver can estimate the environmental conditions at different locations and provide sensing services for the users, e.g., anomaly detection and localization. 

Moreover, as shown in Fig.~\ref{fig: system overview}~(b), the heights of the Rx antennas and the Tx antenna array are set symmetrical with respect to the height of the {\mms} arrays.
This is because for the wireless signals with wavelength larger than the spacing interval of {\mmsu}, the {\mms} arrays is equivalent to a uniform medium like a mirror.
Then, the signal reflection angle on the {\mms} arrays is approximately equal to the incident angle.

Furthermore, the multi-antenna {\rfmeasure} is able to move horizontally along the wall to measure the {\mms}s at different locations.
The mobile platform moves along the $y$-axis to scan the {\mms}s at different locations, keeping with measuring distance~$D$.
Specifically, it moves to $N_{\mathrm{loc}}$ different measurement locations, and uses the {\rfmeasure} to measure the reflected signals of {\mms}s at each location.

\subsection{Received Signal Model}
\label{ssec: received signal model}

\begin{figure}[!t]
\center{\includegraphics[width=0.45\linewidth]{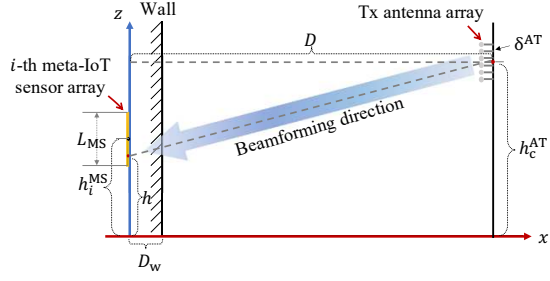}}
    \vspace{-.5em}
    \setlength{\belowcaptionskip}{-1.em} 
    \caption{Vertical section of the meta-IoT system, where the Tx antenna array aims its beam towards the {\mms}s at height $h$ in the $i$-th {\mms} array.}
    \vspace{0.em}
    \label{fig: propagation path illustration}
\end{figure}

In the following, we propose the analytical received signal model for the meta-IoT system, jointly considering the influence of sensing and transmission.
To facilitate analysis, we use the beamforming capability of the Tx antenna array so that the transmitted signal beam is aimed at the $i$-th {\mms} array when measuring it.
Without loss of generality, we assume that the Tx beamforming is performed by the Tx antennas adding additional phases to their transmitted signals, and the beam is aimed at the {\mms}s at height $h$, as shown in Fig.~\ref{fig: propagation path illustration}.

Specifically, we focus on modeling the received signals due to the reflection of meta-IoT sensor arrays, which contain information on environmental conditions. 
By modeling this part of received signals, we facilitate the formulation of the optimization problem for the meta-IoT system's sensing performance.
Other arriving signals at the receiver, such as the reflected signals from the walls, are omitted since they are not sensitive to environmental conditions and cannot contribute to the sensing performance of the meta-IoT system.

Based on the transmission model of metamaterial surfaces proposed and experimentally verified in \cite{tang2019wireless}, the influence of metamaterial particles on normal and oblique incident signals can be approximated by a reflection coefficient multiplied by a normalized radiation pattern factor.
Since the SRR-based meta-IoT sensor has a radiation pattern close to a hemisphere, its normalized radiation pattern can be approximated by $1$, and thus its influence on the signal transmission can be modeled by the reflection coefficient in (\ref{equ: reflection coefficient}).
Thus, the signals transmitted by the $j$-th Tx antenna, reflected by the $i$-th meta-IoT sensor array, and received by the $i$-th Rx antenna can be expressed as
\begin{align}
\label{equ: int channel model}
y_{j,i}(f, \bm d, \bm c, h, D)= &\iint_{\bm x^{\text{M}}\in \mathcal S_i^\mathrm{M}} \sqrt{P} \cdot g_{j}^{\tx}(\bm x^{\tx}_j,\bm x^{\text{M}} , f, h, D) \nonumber\\
& \cdot\gamma(\bm x^{\text{M}}, f, \bm d, \bm c) \cdot g_i^{\rx}(\bm x^{\rx}_i, \bm x^{\text{M}},  f, D) \cdot \mathrm{d}\bm x^{\text{M}}. 
\end{align}

In~(\ref{equ: int channel model}), the surface integral is over the surface of the $i$-th {\mms} array, which is denoted by $\mathcal S_i^\mathrm{M}$, and the differential on the surface is denoted by $\bm x^{\text{M}}$.
In the surface integral, the first term, i.e., $\sqrt{P}$ is the magnitude of the Tx signal with $P$ being the Tx power.
The second term, i.e., $g_{j}^{\tx}(\bm x^{\tx}_j,\bm x^{\text{M}} , f, h, D)$, indicates the channel gain from the $j$-th Tx antenna at $\bm x^{\tx}_j$ to $\bm x^{\text{M}}$.
The third term, i.e., $\gamma(\bm x^{\text{M}}, f, \bm d, \bm c)$ is the reflection coefficient of the meta-IoT sensor at $\bm x^{\text{M}}$, which is modeled based on (\ref{equ: reflection coefficient}).
The reflection coefficient indicates the magnitude ratio and phase shift between the reflected and incident RF signals~\cite{pozar2011microwave}.
Thus, the product of the first three terms represents the reflected RF signals of the meta-IoT sensor at $\bm x^{\text{M}}$, and the magnitude of the scattering RF signals can be calculated by $\sqrt{P} \cdot |g_{j}^{\tx}(\bm x^{\tx}_j,\bm x^{\text{M}} , f, h, D) \cdot \gamma(\bm x^{\text{M}}, f, \bm d, \bm c)|$.
The fourth term, i.e., $g_i^{\rx}(\bm x^{\rx}_i, \bm x^{\text{M}},  f, D)$, indicates the channel gain from $\bm x^{\text{M}}$ to the $i$-th Rx antenna at $\bm x^{\rx}_i$.
Specifically, the second and fourth and terms in~(\ref{equ: int channel model}) can be expressed by 
\begin{align}
\label{equ: g_j^tx}
   g_{j}^{\tx}(\bm x^{\tx}_j,\bm x^{\text{M}} , f, h, D) =  g_{j}^{\tx}(f, \theta_{j,i}^{\tx}) \cdot e^{\iu \cdot \varphi^{\tx}_{j,i}(f, h, D)} \cdot g(\bm x^{\tx}_j,\bm x^{\text{M}} , f, D), 
\end{align}
\vspace{-2.5em}
\begin{align}
\label{equ: g_j^rx}
&g_i^{\rx}(\bm x^{\rx}_i, \bm x^{\text{M}},  f, D) = g_i^{\rx}(f,\theta_i^{\rx})\cdot g(\bm x^{\rx}_i, \bm x^{\text{M}},  f, D),
\end{align}
where $ \varphi_{j,i}^{\tx}(f, h, D)$ is the added phase to the signals of frequency $f$ transmitted by the $j$-th Tx antenna for beamforming when measuring the $i$-th {\mms} array at height $h$ and distance $D$, and $g_{j}^{\tx}(f, \theta_{j,i}^{\tx})$ and $g_{i}^{\rx}(f, \theta_i^{\rx})$ indicate the gains of the $j$-th Tx antenna and the $i$-th Rx antenna, respectively.
Here, $\theta_{j,i}^{\tx}$ denotes the emission angle of the signals from the $j$-th Tx antenna to the $i$-th meta-IoT sensor array, and $\theta_i^{\rx}$ denotes the incident angle of signals from  the $i$-th meta-IoT sensor array to the $i$-th Rx antenna, which can be approximated by
\begin{align}
\label{equ: Tx and Rx angle}
& \theta_{j,i}^{\tx} \!=\! \arctan(\frac{h_j^{\tx}-h_i^{\MS}}{D}),\quad
\theta_{i}^{\rx} \!=\! \arctan(\frac{h_i^{\MS} - h_i^{\rx}}{D}),
\end{align}
where $h_j^{\tx}$, $h_i^{\MS}$, and $h_i^{\rx}$ denote the heights of the $j$-th Tx antenna, the center of the $i$-th meta-IoT sensor array, and the $i$-th Rx antenna, respectively.

Furthermore, $g(\bm x, \bm x^\m, f,  D)$ in (\ref{equ: g_j^tx}) and~(\ref{equ: g_j^rx}) indicates the propagation channel gain from $\bm x$ to $\bm x^\m$, which can be calculated by
\begin{align}
\label{equ: general channel}
 g(\bm x, \bm x^\m, f, D) = 
\underbrace{\frac{ve^{ -\iu \frac{2\pi f}{v}\| \bm x - \bm x^\m \|} }{4\pi f \| \bm x - \bm x^{\m} \|}  }_{\text{\scriptsize Free-space propagation}}\cdot 
\underbrace{e^{-\beta \frac{D_\mathrm{w}}{D}\| \bm x - \bm x^\m \|-\iu \frac{2\pi f D_\mathrm{w}}{vD}(n_\mathrm{w} - 1)\cdot \| \bm x - \bm x^\m\| }}_{\text{\scriptsize In-wall propagation}}, 
\end{align}
where $v$ denotes the speed of light.

In~(\ref{equ: general channel}), the first term is derived from the signal propagation in free-space, and the second term is the factor accounting for the in-wall propagation.
Besides, $\beta$, $n_\w$, and $D_\w$ denote the attenuation factor, the relative refraction index, and the thickness of the wall, respectively.
Moreover, in~(\ref{equ: g_j^tx}), $\varphi_{j,i}^{\tx}(f, h, D)$ aligns the phases of the transmitted signals of the $N_{\AT}$ antennas in the direction towards the $i$-th {\mms} array at height~$h$.

Nevertheless, due to the integration in~(\ref{equ: int channel model}), the channel model is hard to calculate, which results in difficulty when designing the {\mmsd} and setting the added phase for beamforming.
To handle this issue, in the following, we provide computationally efficient models of received signals considering the cases where the numbers of {\mms}s are small and large, respectively.

\subsubsection{Small number of {\mms}s case}
As the size of each {\mms} is on the sub-wavelength scale, the variation of the wave propagation distance and arriving phase across the region of a {\mms} is negligible~\cite{Lu2021Communicating}.
Therefore, we can simplify the integration in~(\ref{equ: int channel model}) by discretizing it into a summation, which can be expressed as
\begin{align}
\label{equ: small number channel model}
 y_{j,i}^{\text{small}}(f, \bm d, \bm c, h, D) \approx & \sqrt{P} \cdot \sum_{m\in \mathcal N_i^\m} g(\bm x^{\tx}_j,\bm x_m^{\text{M}} , f, D) \\
&\cdot e^{\iu \varphi^{\tx}_{j,i}(f, h, D)} \cdot \gamma(f, \bm d, \bm c) \cdot g(\bm x^{\rx}_i, \bm x_m^{\text{M}}, f, D) \cdot A, \nonumber
\end{align}
where $A = N_\T l_\s^2$ is the area of a {\mms}
and $\mathcal N_i^\m$ denotes index set of the {\mms}s in the $i$-th array,
$\bm x_m^{\text{M}}$ is the coordinate of the center of the $m$-th {\mms}.

Equation~(\ref{equ: small number channel model}) is accurate and computational efficient when the number of {\mms}s is small. 
Nevertheless, (\ref{equ: small number channel model}) is not fit the case whose number of {\mms}s is large as the summation still incurs high computation complexity.

\subsubsection{Large number of {\mms}s case}

As expected in 6G communications, the sensors should be massively deployed to support ubiquitous sensing.
Therefore, it is both necessary and important to obtain a computationally efficient received signal model for cases where large numbers of {\mms}s are deployed.
Thanks to the effective medium property of the {\mm}s, we can consider the {\mms} array with a large number of {\mms}s as a uniform reflective surface.
Thus, the reflection of the wireless signals on a {\mms} array is similar to the reflection of light on a uniform surface.
Therefore, we can obtain a much simplified received signal model by the following proposition.

\begin{proposition}
\label{prop: channel model}
Assume that the $i$-th {\mms} array is electrically large, i.e., both its length and width are sufficiently larger than the wavelength $\lambda=v/f$ of the incident signals.
Then, the received signal of the $i$-th Rx antenna for the signals transmitted by the $j$-th Tx antenna and reflected by the $i$-th {\mms} array can be approximated by
\begin{align}
\label{equ: prop channel model equ 1}
y_{j,i}^{\text{large}}(f, \bm d, \bm c, h,  D)   \approx & \sqrt{P}\cdot g_{j}^{\tx}(f, \theta_{j,i}^{\tx})\cdot e^{\iu \varphi_{j,i}^{\tx}(f, h, D)}  \\
&\cdot g_i^{\rx}(f,\theta_i^{\rx}) \cdot\chi(f,\bm d,\bm c, D) \cdot g_{\text{sr}}({\bm x}^{\rx}_i, {\bm x}^{\tx}_j, f, D),  \nonumber
\end{align}
where $\chi(f, \bm d, \bm c, D)$ is a coefficient due to the reflection on the {\mms} array, and $g_{\text{sr}}({\bm x}^{\rx}_i, {\bm x}^{\tx}_j, f, D)$ is the gain of the specular reflection path from ${\bm x}^{\tx}_j$ to $\bm x^\rx_i$, i.e.,
\begin{align}
\label{equ: g sr}
g_{\text{sr}}({\bm x}^{\rx}_i, \ & {\bm x}^{\tx}_j, f, D) = 
\frac{ve^{ -\iu \frac{2\pi f}{v}\| \tilde{\bm x}^{\rx}_i - {\bm x}^{\tx}_j \|} }{4\pi f \| \tilde{\bm x}^{\rx}_i - {\bm x}^{\tx}_j \|} \cdot \ e^{-\beta \frac{D_\mathrm{w}}{D}\| \tilde{\bm x}^{\rx}_i  - {\bm x}^{\tx}_j \|-\iu \frac{2\pi f D_\mathrm{w}}{vD}(n_\mathrm{w} - 1)\cdot \| \tilde{\bm x}^{\rx}_i - {\bm x}^{\tx}_j\| }, 
\end{align}
where $\tilde{\bm x}^{\rx}_i$ is the symmetric point of $\bm x^{\rx}_i$ with respect to the surface of {\mms} arrays.
Besides, $\chi(f, \bm d, \bm c, D)$ can be calculated by
 \beq
 \label{equ: prop A chi}
 \chi(f, \bm d, \bm c, D) \!=\! \frac{Dv^2   \gamma(f, \bm d, \bm c)}{2\beta {D_\w} v f \!+\! 4\pi f^2\iu ( D \!+\! {D_\w} (n_\w \!-\!1 ))}.
 \eeq
\end{proposition}
\begin{IEEEproof}
See Appendix~\ref{appx: derivation of large number channel model}.	
\end{IEEEproof}

As the number of {\mms}s in the considered system is large, we adopt the received signal model~(\ref{equ: prop channel model equ 1}) in this paper.
Without loss of generality, we assume that the received signals are measured at $N_\F$ frequencies, which is denoted by set $\mathcal F$.
Consequently, based on~(\ref{equ: prop channel model equ 1}) the received signal vector of the $i$-th Rx antenna is can be denoted by 
\begin{align}
\label{equ: y_i equals sum}
\bm y_i(\bm d,  \bm c, \ & h, D) = \sum_{j=1}^{N_{\AT}}(y^{\text{large}}_{j,i}(f_1, \bm d, \bm c, h, D),\dots  y_{j,i}^{\text{large}}(f_{N_\F}, \bm d, \bm c, h, D)).
\end{align}

Moreover, based on~(\ref{equ: g sr}), to align the phase of the received signals when different Tx antennas are transmitting, the added phase for beamforming in (\ref{equ: small number channel model}) can be set to
\begin{align}
&\varphi_{j,i}^{\tx}(f, h, D) = (\frac{2\pi f}{v}+\frac{2\pi f D_{\w}}{vD}(n_{\w}-1) )\cdot (\| \bm x_j^{\tx} - \tilde{\bm x}_i^{\rx}\| - \| \bm x_1^{\tx} - \tilde{\bm x}_i^{\rx}\|).
\end{align}
Given that $D\gg \delta^{\AT}$ and $D\gg D_{\w}$, the above equation can be approximated by
\begin{align}
\label{equ: setting varphi}
\varphi_{j,i}^{\tx}(f, h, D) \approx & -\frac{(j-1)\cdot 2\pi f \delta^{\AT}}{v}\cdot \frac{h-h_\cen^\AT}{\sqrt{D^2+(h-h_\cen^\AT)^2}}, h\in[h_i^{\MS} - L_{\MS}/2, h_i^{\MS} + L_{\MS}/2],
\end{align}
where $h_\cen^\AT$ denotes the center height of the Tx antenna array, 
$\delta^{\AT}$ is the interval between adjacent Tx antennas,
and $L_{\MS}=N_{\MS}\cdot l_\s$ denotes the vertical length of a {\mms} array with $N_{\MS}$ being the number of sensors in a column.

\textbf{Computational complexity analysis}: The complexity of calculating the $N_F$-dimensional received signal vectors of the $N_{\mathrm{AR}}$ meta-IoT sensor arrays by (\ref{equ: y_i equals sum}) can be analyzed as follows.
To calculate the $N_F$-dimensional received signal vectors of the $N_{\mathrm{AR}}$ meta-IoT sensor arrays, Eq.~(\ref{equ: prop channel model equ 1}) needs to be calculated for $N_F N_{\mathrm{AT}} N_{\mathrm{AR}}$ times.
As for Eq.~(\ref{equ: prop channel model equ 1}), its computational complexity is in proportion to that of reflection coefficient $\gamma(f, \bm d, \bm c)$, which is calculated by Eqs. (\ref{equ: total impedance}) and~(\ref{equ: reflection coefficient}).
Based on Eq. (\ref{equ: total impedance}), the calculation of $\gamma(f, \bm d, \bm c)$ has the complexity of $\mathcal O(N_T)$.
Therefore, in summary, the computational complexity of all the received signal vectors is $\mathcal O(N_T N_F N_{\mathrm{AT}} N_{\mathrm{AR}})$.

In contrast, if the received signal model for the small number of meta-IoT sensors case is adopted, then based on (\ref{equ: small number channel model}) and the above analyses, the computational complexity of all the received signal vectors is $\mathcal O(N_{M}N_T N_F N_{\mathrm{AT}} N_{\mathrm{AR}})$, where $N_{M}$ denotes the number of meta-IoT sensors in a meta-IoT sensor array.

It can be observed that the computational complexity of using the received signal model for the small number of sensors case is $N_{M}$ times higher than using that for the large number of sensors case.
Since $N_{M}$ is large in the considered scenarios, then using Eq.~(\ref{equ: prop channel model equ 1}) rather than (\ref{equ: small number channel model}) for the calculation of received signals can efficiently reduce the computational complexity.

\section{Joint Sensing and Transmission Optimization for {\MmsD}}
\label{sec: joint sensing and transmission optimization}

In this section, we first formulate the meta-IoT structure optimization problem of the meta-IoT system in Section~\ref{ssec: problem formulation}. For generality, we consider the objective to maximize the discernibility of received signals under different environmental conditions, which is suitable for various sensing applications. Then, an efficient meta-IoT structure optimization algorithm is proposed in Section~\ref{ssec: algorithm design}.

\subsection{Problem Formulation}
\label{ssec: problem formulation}

Before formulating the optimization problem of the {\mmsd}, we first specify the feature vectors obtained from the received signals, which are used to evaluate the environmental conditions.
For the sake of feasibility of the meta-IoT system, we take the amplitude of the received signals to obtain the sensing results since the measurement of signal amplitude is generally easier than that of the signal phase, which requires less sophisticated equipments.
Besides, to magnify the difference between small signal amplitudes, we use decibel measure instead of decimal measure.
Moreover, for each {\mms} array, multiple heights are taken into consideration in the Tx beamforming, which is denoted by a \emph{height displacement set} $\mathcal S_{\mathrm{dH}} =\{\Delta h_m = -L_{\mathrm{MS}}/2 + (m-1)\cdot L_{\mathrm{MS}}/ N_{\mathrm{dH}}| m\in[1,N_{\mathrm{dH}}] \} $ with $|\mathcal S_{\mathrm{dH}}| = N_{\mathrm{dH}}$.
Furthermore, assume that the wireless transceiver has $N_{\mathrm{MD}}$ different potential measuring distances, which is denoted by $D_{{}q}$~($q\in[1,N_{\mathrm{MD}}]$).
Therefore, given environmental condition $\bm c$ and a certain measurement location, the \emph{feature vector} to obtain the sensing result for the $i$-th {\mms} array can be expressed as
\begin{align}
\label{equ: feature vector}
\bm p_{i,m}( \bm c; \bm d, D_{{}q}) = & 10\log(|\bm y_i(\bm d, \bm c, h_i^{\mathrm{MS}} + \Delta h_m, D_{{}q})|),   i\!\in\![1, N_{\mathrm{AR}}], m\!\in\![1, N_{\mathrm{dH}}], q\!\in\![1,N_{\mathrm{MD}}].
\end{align}

Then, based on (\ref{equ: feature vector}), the formulation of {\mmsd} optimization can be described as follow.
Without loss of generality, we consider the sensing algorithm to infer {\senscond}s from the received signals as a general classification algorithm.
To ensure classification accuracy, it demands the received reflected signals under different {\senscond}s to be as \emph{discernible} as possible.

Consider $\bm g^{\bm w}$ as a general classification function, which infers the {\envircondvec} corresponding to a feature vector, and thus we need to maximize the \emph{discernibility} of the feature vectors for different environmental conditions in order to maximize the classification performance.
One of the widely adopted discernibility measures is the  Euclidean distance, which is used to evaluate the distance between two vectors and indicates how much two vectors can be discerned from each other~\cite{goodfellow2016deep}.
Intuitively, a smaller Euclidean distance indicates a larger discernibility of the two vectors.
Specifically, the error probability to judge between two environmental conditions is dependent on the Euclidean distance between their corresponding feature vectors, which can be calculated based on Proposition~\ref{prop: distinguish ability}.

\begin{proposition}
\label{prop: distinguish ability}
Assume that the maximum likelihood criterion is adopted to judge between $\bm c_{l}$, $\bm c_{l'}$ based on their corresponding feature vectors $\bm p_{i,m}(\bm c_l; \bm d, D_{{}q})$ and $\bm p_{i,m}(\bm c_{l'};\bm d, D_{{}q})$.
Then, the error probability to judge $\bm c_{l}$ erroneously as $\bm c_{l'}$ is in inverse relation to Euclidean distance $\|\bm p_{i,m}(\bm c_{l}; \bm d, D_{{}q}) - \bm p_{i,m}(\bm c_{l'}; \bm d, D_{{}q})\|_2$ and can be calculated by 
\begin{align}
\label{equ: pr err}
\Pr^{\mathrm{err}}&(\bm c_{l'}|\bm c_l) = {\frac{1}{2}-\frac{1}{2}\cdot \erf \big({\|\bm p_{i,m}(\bm c_{l}; \bm d, D_{{}q}) - \bm p_{i,m}(\bm c_{l'}; \bm d, D_{{}q})\|_2 \over 2\sqrt{2\sigma_\mathrm{M}^2}} \big) },
\end{align}
where $\erf(\cdot)$ denotes the \emph{error function}~\cite{McDonough_SIGNAL}{}, and $\sigma^2_\mathrm{M}$ is the variance of the measurement noise following a normal distribution.
\end{proposition}
\begin{IEEEproof}
See the proof of Proposition~1 in~\cite{Hu2021Meta-IoTTWC}. 
\end{IEEEproof}

As shown in Proposition~\ref{prop: distinguish ability}, the error probability to judge between two environmental conditions is dependent on the Euclidean distance between feature vectors corresponding to the two environmental conditions as well as the variance of measurement noise, i.e., $\sigma_\mathrm{M}^2$.
Nevertheless, the variance of measurement noise is hard to estimate and varies for different systems and operating conditions.
Thus, for generality, we evaluate the discernibility by the Euclidean distance.
Therefore, based on Proposition~\ref{prop: distinguish ability} and~(\ref{equ: prop channel model equ 1})$\sim$(\ref{equ: feature vector}), the joint sensing and transmission optimization problem of the {\mmsd} can be formulated as follows:
\manuallabel{opt: sensor design optimization}{P1}
\begin{subequations}
\begin{align}
\label{opt P1 obj func}
&\text{(\ref{opt: sensor design optimization}):}  \max_{\bm d} ~%
\Psi(\bm d) \!= \!
\frac{1}{N_\mathrm{C}N_{\mathrm{AR}}N_{\mathrm{dH}}N_{\mathrm{MD}}}\!\cdot \!\Big(
\sum_{l=1}^{N_{\mathrm{C}}}
\sum_{i=1}^{N_{\mathrm{AR}}}
\sum_{m=1}^{N_{\mathrm{dH}}} 
\sum_{q=1}^{N_{\mathrm{MD}}}
 \|\bm p_{i,m}(\bm c_{l}; \bm d, D_{{}q}) - \bm p_{i,m}(\bm c_{l'}; \bm d, D_{{}q})\|_2
\Big),  \\
\label{constraint: p1-1}
&s.t.~~ y_{j,i}^{\text{large}}(f, \bm d, \bm c, h, D_q) = \gamma(f, \bm d, \bm c_l)  \tilde{g}_{i,j}(f, h, D_q), \\
\label{constraint: p1-2}
& \qquad \tilde{g}_{i,j}(f, h, D_q) \!=\! \frac{g_{j}^{\tx}(f, \theta_{j,i}^{\tx}) e^{\iu \varphi_{j,i}^{\tx}(f, h, D_q)}  g_i^{\rx}(f,\theta_i^{\rx})}{2\beta {D_\w} v f+ 4\pi f^2\iu ( D_q + {D_\w} (n_\w -1 ))} \cdot \sqrt{P} \cdot D_qv^2 \cdot g_{\text{sr}}({\bm x}^{\rx}_i, {\bm x}^{\tx}_j, f, D_q) \\
\label{constraint: p1-4}
& \qquad \bm d \in \mathcal D_\mathrm{A}, \\
& \qquad \text{(\ref{equ: g sr}), (\ref{equ: y_i equals sum}), (\ref{equ: setting varphi}), (\ref{equ: feature vector})}. \nonumber
\end{align}
\end{subequations}

In~(P1), the physical meaning of objective function $\Psi(\bm d)$ is the average discernibility of the feature vectors under different environmental conditions in terms of their Euclidean distance.
Constraint~(\ref{constraint: p1-1}) indicates that the received signals are determined by reflection coefficient of {\mms}s $\gamma(f, \bm d, \bm c)$ and an equivalent channel gain denoted by $\tilde{g}_{i,j}(f, h, D)$, which can be calculated by~(\ref{constraint: p1-2}) based on~(\ref{equ: prop channel model equ 1}) and~(\ref{equ: prop A chi}). It can be observed in~(\ref{constraint: p1-1}) that $\bm d$ influences the reflection coefficients for different {\senscond} $\bm c$, which represents the sensitivity of meta-IoT sensors, as well as the received signals given each $\bm c$, which represents the transmission of RF signals.
Therefore, {\mmsd} $\bm d$ jointly influences the sensing and transmission of meta-IoT systems.

Moreover, constraint~(\ref{constraint: p1-4}) means $\bm d$ takes value within an available {\mmsd} set denoted by $\mathcal D_\mathrm{A} = \{\bm d =(d_1,...d_i,...d_{N_T})| d_i \in [d_{\mathrm{lb}}, d_{\mathrm{ub}}], d_i \in \mathbb R, \forall i\in[1,N_T] \}$.
Specifically, $d_{\mathrm{lb}}$ and $d_{\mathrm{ub}}$ are the lower and upper bounds of the gap widths, respectively, which are acquired by using preliminary simulation trials to ensure the resonant absorption peaks of the {\mms} within frequency range $[f_{\mathrm{lb}}, f_{\mathrm{ub}}]$. 

Finally, the last row of (P1) indicates that the feature vectors of the meta-IoT system follow the channel model described by~(\ref{equ: g sr}),~(\ref{equ: y_i equals sum}),~(\ref{equ: setting varphi}) and~(\ref{equ: feature vector}). 
Based on~(\ref{equ: y_i equals sum}) and~(\ref{equ: feature vector}), the feature vectors in objective function $\Psi(\bm d)$ are composed of the received signals defined by~(\ref{constraint: p1-1}).
Thus, the average discernibility is under the joint influence of $\bm d$ on sensing and transmission.
Therefore, by maximizing the average discernibility with respect to $\bm d$, both the sensing and transmission of {\mms}s are optimized.

\subsection{Algorithm Design}
\label{ssec: algorithm design}
In solving (P1), the challenge lies in that the unacceptable high computational complexity to evaluate the $\Psi(\bm d)$.
This is majorly due to calculating the precise reflection coefficient function, which generally has no closed-form expression and requires highly time-consuming full-wave simulation to evaluate. 
To handle this challenge, we first obtain a fitted function to substitute the precise reflection coefficient function.
For $n\in[1,N_F]$, the fitted function is based on the analytical model derived in Section~\ref{ssec: ref coeff model}, which can be expressed as
\begin{equation}
\label{equ: rcs fitting function with polymer}
{\gamma}(f_n, \bm d, \bm c) = \hat{\gamma}(f_n, \bm d,\bm c) + \xi_{n}(\bm d, \bm c),
\end{equation}
where $\xi_{n}(\bm d, \bm c)$ is an interpolated function of $\bm d$ used to eliminate the error between the model prediction and the precise full-wave simulation results. 

\begin{algorithm}[t]
\small
\caption{{\Mmsd} optimization algorithm}
\label{alg: design optimization algorithm}
\hspace*{0.02in} {\bf Input:} 
Set of {\senscond} vectors $\mathcal C$, $|\mathcal C| = N_C$;
Set of frequency points $\mathcal F$, $|\mathcal F| = N_F$;
Set of available {\mmsd}s, $\mathcal D_\mathrm{A}$;
Set of sampled {\mmsd}s with precise reflection coefficient functions, $\hat{\mathcal D}_\mathrm{A}\subset \mathcal D_\mathrm{A}$;
Distance threshold in the terminal condition for the surrogate algorithm $\upsilon_{\min}$.
\\
\hspace*{0.02in} {\bf Output:} 
Optimal {\mmsd} $\bm d^*$.
\begin{algorithmic} [1]

\State Fit $\gamma(f, \bm d, \bm c)$ with 
$\hat{\gamma}(f, \bm d, \bm c)$ in~(\ref{equ: rcs fitting function with polymer}) given $\bm d\in\hat{\mathcal D}_\mathrm{A}$, $\bm c \in\mathcal C$.

\State Generate a random set of samples within $ \mathcal D_\mathrm{A}$, and denote it by $\mathcal S_{\mathrm{obj}}$.

\State Evaluate the corresponding objective values in (\ref{opt: sensor design optimization}) of the samples in $\mathcal S_{\mathrm{obj}}$.
 
\State Construct a surrogate objective function by interpolating the samples in $\mathcal S_{\mathrm{obj}}$ with a radial basis function~\cite{Buhmann_Radial}.

\State Generate a new set of random samples within $\mathcal D_\mathrm{A}$, and denote it by $\mathcal S_{\mathrm{sur}}$.

\State Evaluate elements in $\mathcal S_{\mathrm{sur}}$ by the surrogate function and select the best point to be the candidate point, which is included into $\mathcal S_{\mathrm{obj}}$. 

\State If all the samples in $\mathcal S_{\mathrm{sur}}$ are within the $\upsilon_{\min}$ distance from those in $\mathcal S_{\mathrm{obj}}$, go to Step $8$; otherwise, go to Step~$4$.

\State \Return the element in $\mathcal S_{\mathrm{obj}}$ with the current minimum objective function value as $\bm d^*$.
\end{algorithmic}
\end{algorithm}

By using ${\gamma}(f, \bm d, \bm c)$ in (P1), the original optimization problem becomes a more computationally feasible.
Then, as the objective function in (\ref{opt: sensor design optimization}) contains the non-convex function $\xi_{n}(\bm d, \bm c)$ in~(\ref{equ: rcs fitting function with polymer}), (P1) is non-convex, which is hard to solve.
To solve (\ref{opt: sensor design optimization}) efficiently, we can adopt efficient algorithms such as the pattern search algorithm~\cite{Wang2014AGeneral} or the surrogate optimization algorithm as in~\cite{Hu2021Meta-IoTTWC}, which can handle finitely bounded non-convex optimization problems and has a high probability of finding global optimum.
The proposed algorithm for solving~(\ref{opt: sensor design optimization}) is summarized as Algorithm~\ref{alg: design optimization algorithm}.
The resulting optimal {\mmsd} vector is denoted by $\bm d^*$.

\section{Unsupervised Sensing Algorithm Design for Meta-IoT System}
\label{sec: anomal detect and locate mod}

In this section, we handle the sensing algorithm design in the meta-IoT system. As an example, we apply the meta-IoT system to anomaly detection and localization, which is one of the most important sensing applications in 6G~\cite{Han2021Anomaly}.

We first state the anomaly detection and localization problem in designing the sensing algorithm in Section~\ref{ssec: problem statement} and then propose an unsupervised sensing algorithm to solve it efficiently Section~\ref{ssec: unsupervised alg design}.

\subsection{Problem Statement}
\label{ssec: problem statement}

The anomaly detection and localization problem is stated as follows:
Given a historical time-series data set $\mathcal T$ which is composed of $N_F$-dim feature vectors collected over time $T$, and assuming that no anomaly exists during the data collection, the sensing algorithm aims to achieve two goals:
\begin{itemize}[leftmargin=1.5em]
\item \textbf{Detecting Anomaly}: Detecting a happened environmental anomaly at certain time steps after $T$ and interpreting the environmental anomaly severity qualitatively.
\item \textbf{Locating Anomaly}: Locating the position where the environmental anomaly took place.
\end{itemize}

For a sensing algorithm to achieve the above goals, it is important that the following three challenges are handled.

\begin{enumerate}[leftmargin=1.5em]
\item As {\senscond}s are indicated by the frequency responses of the {\mms}s, the algorithm needs to extract effective spectrum features of the received signals.
\item The algorithm needs to be robust with the impact of multi-path and noises, which generally exist in wireless systems.
\item The sensing algorithm needs to detect and locate anomalies by learning from an unlabeled data set. This is because, the meta-IoT system is designed to collect training data in an automatic manner by itself, and no auxiliary systems are adopted to inform it of the current environmental conditions.
\end{enumerate}

One of the solutions is to consider the problem as a traditional channel estimation problem, where the sensing algorithm recovers the intact reflection coefficients of {\mms}s to detect and locate anomalies.
Such methods are proposed for intelligent meta-material surfaces and RFID backscatters based on deep supervised learning~\cite{Liu2022Deep, Liu2021Deep}.

Nevertheless, in contrast to intelligent meta-material surfaces and RFID backscatters, the meta-IoT sensors are passive and chipless, and their reflection coefficients are determined by unknown environmental conditions. 
This indicates that in the meta-IoT systems, the sensors cannot configure themselves to predefined states and provide the training examples, e.g., pilots, which are needed in the supervised learning.

\begin{figure}[!t]
\center{\includegraphics[width=0.6\linewidth]{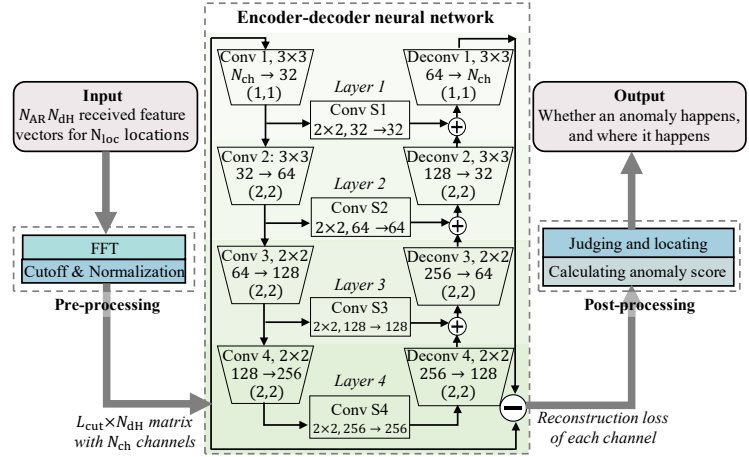}}
	\vspace{-0.5em}
	\setlength{\belowcaptionskip}{-1.em} 	\setlength{\abovecaptionskip}{-.1em}
	\caption{
	Data-flow diagram of the proposed sensing algorithm.}
	\vspace{0.em}
	\label{fig: neural net}
\end{figure}

To handle the above challenges, we propose a deep unsupervised learning sensing algorithm, which is composed of a pre-processing part, a convolutional encoder-decoder neural network, and a post-processing part.
Firstly, in the pre-processing part, we employ the fast Fourier transformation~(FFT) to extract spectrum features from the reflected signals of each {\mms} array.
Besides, the spectrum features obtained by FFT are further truncated, which reduces the rapid fluctuations in the spectrum features caused by multi-path and noises.
Secondly, we use the convolutional encoder-decoder neural network which is efficient in noise reduction to further extract the principal information of the input features~\cite{Eargle1992}.
Due to the pre-processing part and the encoder-decoder neural network, the proposed sensing algorithm can also be robust to possible external interference from other wireless systems\footnote{
If the extern interference is large at the frequencies of the meta-IoT sensors' resonant absorption peaks, then the performance of the sensing algorithm may decline. To handle this issue, Rx antennas with higher directionality can be used, whose main lobes are pointed at meta-IoT sensor arrays to reduce external interference.}.

The neural network is trained by using the unsupervised learning technique with unlabeled data.
In contrast to supervised learning where labeled data are in need, unsupervised learning is able to learn the similarity and differences of the measurement data within unlabeled training data set~\cite{hinton1999unsupervised}.
By calculating the difference between the measurement data, a sensing algorithm based on unsupervised learning can discriminate abnormal data from normal ones.

Besides, as the neural network learns directly from the collected data, it learns the relationship between the environmental conditions and the received signals automatically without relying on specific wireless propagation channels, such as dominating line-of-sight channels. 
Thus, though the multi-path effect makes received signals specific to wireless environments, it does not influence the environmental condition recognition of the proposed sensing algorithm. 
Thirdly, in the post-processing part, we compare the input and output features of the encoder-decoder neural network and derive the occurrence and location of the environmental anomalies.

\subsection{Algorithm Design}
\label{ssec: unsupervised alg design}
To solve the problem stated above, the following unsupervised sensing algorithm is proposed.
The proposed algorithm takes in the feature vectors obtained from the received signals at a certain measuring distance $D$, and outputs indicators of the anomaly detection and localization.
The complete data-flow diagram of the proposed algorithm is shown in Fig.~\ref{fig: neural net}, and the blocks in the diagram are introduced sequentially as follows.

\vspace{0.3em}
\textbf{Input}:
The input data to this block is the measurement data of time $t$.
Specifically, for a measurement indexed by time $t$, the input data set contains the $N_{\mathrm{AR}}N_{\mathrm{dH}}$ feature vectors at each location $k$~($\forall k\in[1, N_{\mathrm{loc}}]$), which can be expressed as
\beq
\mathcal D^t \!=\! \{ \bm p_{i,m}^{k,t}\ | i\in[1,N_{\mathrm{AR}}], m\in[1, N_{\mathrm{dH}}], k\in[1, N_{\mathrm{loc}}]\}.
\eeq
Here, $\bm p_{i,m}^{k,t}$ denotes the feature vector defined in~(\ref{equ: feature vector}), which is collected at time $t$ and location $k$ with measuring distance $D$, given optimal {\mmsd} $\bm d^*$.

\vspace{0.3em}
\textbf{FFT}:
In order to extract the spectrum feature from $\bm p_{i,m}^{k,t}$, we propose to handle the input feature vectors by using FFT.
Specifically, in the \emph{FFT} block, it applies FFT on each vector $\bm p_{i,m}^{k,t}$ in set ${\mathcal D}^t$.
The resulting vector of applying FFT on $\bm p_{i,m}^{k,t}$ is denoted by $\bm P_{i,m}^{k,t}$.

\vspace{0.3em}
\textbf{Cut-off \& Normalization}:
In this block, the results obtained by the FFT block are firstly truncated to length $L_{\mathrm{cut}}$, i.e., $
\tilde{\bm P}_{i,m}^{k,t} = \big( (\bm P_{i,m}^{k, t})_{2},\dots, (\bm P_{i,m}^{k, t})_{L_{\mathrm{cut}+1}}  \big)$.
Here, we remove the first element of $\bm P_{i,m}^{k, t}$ since it indicates an average bias of the received signals, which does not contain the shape feature of the received signals' spectrum.
Moreover, the \emph{min-max normalization} is applied on the truncated vectors, which helps the training of the neural network converge faster.
The min-max normalization can be expressed as
\beq
\hat{\bm P}_{i,m}^{k,t} = \frac{\tilde{\bm P}_{i,m}^{k,t} - \min_t(\tilde{\bm P}_{i,m}^{k,t})}{\max_t(\tilde{\bm P}_{i,m}^{k,t}) - \min_t(\tilde{\bm P}_{i,m}^{k,t})}.
\eeq

Finally, for measurement time $t$, the normalized data is arranged as a $N_{\mathrm{ch}}$-channel $L_{\mathrm{cut}} \times N_\mathrm{dH} $ matrix, which is denoted by $\mathcal M^t$.
Here, $N_{\mathrm{ch}} = N_{\mathrm{AR}}N_{\mathrm{loc}}$ denotes the number of channels.
Specifically, the matrix in the $g_{k,i}$-th~($g_{k,i}=(k-1)\cdot N_{\mathrm{AR}} + i$, $k\in[1, N_{\mathrm{loc}}]$, $i\in[1, N_{\mathrm{AR}}]$) channel can be expressed as $\bm M^t_{k, i} = \big(  \hat{\bm P}_{i,1}^{k,t}, \dots, \hat{\bm P}_{i,N_\mathrm{dH}}^{k,t} \big)$.

\vspace{0.3em}
\textbf{\CoreNetName}:
Data $\mathcal M^t$ is passed to the {\corenetname}. 
The {\corenetname} is composed of multiple connected encoder-decoder layers, the number of which is referred to as the \emph{depth} of the {\corenetname}.
To facilitate description, we describe an encoder-decoder neural network with depth $4$ as used in~\cite{Zhang2019ADeep}, which is shown in Fig.~\ref{fig: neural net} and used in the implemented prototype.
The network takes $\mathcal M^t$ as an input, and outputs a \emph{difference matrix} between $\mathcal M^t$ and a reconstructed $\mathcal M^t$ after being encoded and decoded, which is denoted by $\tilde{\mathcal M}^{t}$.

Specifically, the encoder-decoder layers are composed of convolutional neural networks, which can extract feature maps as the output from input data efficiently~\cite{goodfellow2016deep}.
As shown in Fig.~\ref{fig: neural net}, the four encoder convolutional layers are denoted by Conv 1-Conv 4, whose kernel sizes, input channel sizes, and output channel sizes are given\footnote{
The hyperparameters, including the depth of encoder-decoder neural networks, the kernel sizes, and input channel sizes, are selected following those in~\cite{Zhang2019ADeep}, which perform effectively in our experimental evaluations. To further optimize these hyperparameters, existing methods introduced in~\cite{Yang2020OnHyperparameter} can be used.}.
For example, Conv 1 has a kernel size of $3\times 3$, a stride size of $(1,1)$, an input channel size $N_{\mathrm{ch}}= N_{\mathrm{AR}}N_{\mathrm{loc}}$, and an output channel size of $32$.
The results of four convolutional layers are input to four \emph{state convolutional layers}, i.e., Conv S1$\sim$S4, whose output sizes equal to their input sizes.
The four state convolutional layers provide additional capabilities for extracting the state information from feature maps at different levels.

The outputs of four state convolutional layers contain the spectrum features of the received signals at different measuring heights and the mutual dependence information of different measurement locations in different depths of encoding.
Combining the outputs of the four state convolutional layers, four de-convolutional layers are used to reconstruct the original input, i.e., $\mathcal M^t$, which are denoted by Deconv 1-Deconv 4 in Fig.~\ref{fig: neural net}.
Each decoder layer incorporates the output of a lower decoder layer and the output of a state convolutional layer, and collectively they output the reconstruction of $\mathcal M^t$, i.e., $\tilde{\mathcal M}^t$.
By using this specific structure, the {\corenetname} is effective to infer the substantial information and correlation contained in the input data and can generate accurate reconstructions while suppressing the influence of noises.

Moreover, the reconstruction loss is calculated by differencing $\mathcal M^t$  and $\tilde{\mathcal M}^t$.
Specifically, for the $g_{k,i}$-th channel, the reconstruction loss is calculated by 
\beq
\label{equ: reconstruction loss of channel}
\mathcal L_{k,i} = \| \bm M^t_{k,i} - \tilde{\bm M}^t_{k,i} \|_2^2,
\eeq
where $\|\cdot\|_2$ indicates the $l_2$-norm. 
Then, the \emph{mean reconstruction loss} can be defined by
\beq
\mathcal L_{\mathrm{mean}} = \frac{1}{N_\mathrm{dH}N_{\mathrm{AR}}L_{\mathrm{cut}}N_{\mathrm{loc}}}\sum_{k=1}^{N_{\mathrm{loc}}}\sum_{i=1}^{N_{\mathrm{AR}}}\mathcal L_{k, i}.
\eeq

Using a collected data set $\{\mathcal M^t\}_{t}$, we train the encoder-decoder neural network to minimize the mean reconstruction loss by using the \emph{Adam optimization algorithm}, which is an efficient algorithm for training neural networks~\cite{goodfellow2016deep}.
The training is unsupervised, where no label is provided for the data.
Intuitively, when the trained encoder-decoder neural network is input with new data collected under normal conditions, it has low reconstruction loss of each channel, since the input data is essentially the same as those in the training data set except for some random noises. 
On the other hand, when an anomaly happens, the new data is essentially different from those in the training data set, which leads to large reconstruction losses.

\vspace{0.3em}
\textbf{Calculating Anomaly Score}:
The reconstruction loss of each channel is input into this block.
As described in the previous block, the reconstruction losses can be considered as indicators for anomalies.
Therefore, based on $\mathcal L_{k,i}$ in~(\ref{equ: reconstruction loss of channel}), we propose to evaluate the anomalous degree of the {\senscond}s surrounding the $i$-th {\mms} array at the $k$-th location by using an \emph{anomaly score} calculated by
\beq
\label{equ: anomaly score}
\tau_{k,i}^t = {{\mathcal L}^t_{k,i}}/{\bar{\mathcal L}_{k,i}}.
\eeq
Here, $\bar{\mathcal L}_{k, i}$ denotes the average reconstruction loss of the $g_{k,i}$-th channel given data collected before time $T$, which are stored in the {\rfmeasure}.

\vspace{0.3em}
\textbf{Judging and Locating}:
The anomaly scores are sent to the \emph{judging and locating} block, and the \emph{detection indicator} of an anomaly is calculated by
\beq
\label{equ: occurrence of an anomaly}
 I_{\mathrm{ano}}^t = \begin{cases}
 	1,\quad \text{if $\mathit\Gamma = \frac{\sum_{k=1}^{N_{\mathrm{loc}}}\sum_{i=1}^{N_{\mathrm{AR}}}\tau_{k,i}^t}{N_{\mathrm{loc}}N_{\mathrm{AR}}} > \mathit{\Gamma}_{\mathrm{th}}$},\\
 	0, \quad \text{otherwise}.
 \end{cases}
\eeq
Here, $\mathit{\Gamma}_{\mathrm{\mathrm{th}}} $ is the threshold for anomaly detection, which is selected based on average $\mathit{\Gamma}$ value before $T$.

Given an anomaly is detected, the location of the anomaly is determined as follows.
In this paper, we assume that only one anomaly may occur at a time, and the vertical and horizontal locations of the anomaly can be obtained by 
\beq
\label{equ: location of anomaly}
I_{\mathrm{ver}}^t = \argmax_{i \in [1, N_{\mathrm{AR}}]} \sum_{k=1} ^{N_{\mathrm{loc}}} \tau_{k,i}^t,\quad I_{\mathrm{hor}}^t = \argmax_{k\in[1,N_{\mathrm{loc}}]} \tau_{k, I_{\mathrm{ver}}^t}^t.
\eeq
In~(\ref{equ: location of anomaly}), $I_{\mathrm{ver}}^t$ denotes the vertical location, i.e., height, of the detected anomaly, which is indicated by the index of {\mms} array.
Besides, $I_{\mathrm{hor}}^t$ denotes the horizontal location of the detected anomaly, which is indicated by the index of measurement location.

\vspace{0.3em}
\textbf{Output}:
When $t\leq T$, the sensing algorithm uses data $\mathcal D^t$ for training the encoder-decoder neural network to minimize $\mathcal L_{\mathrm{mean}}$.
On the other hand, when $t> T$, the sensing algorithm outputs detection indicator $I_{\mathrm{ano}}^t$ and anomaly location indicators $(I_{\mathrm{ver}}^t, I_{\mathrm{hor}}^t)$ as the sensing results.

\vspace{0.3em}
\emph{\textbf{Computational complexity analysis}}:
In the pre-processing part, based on the computational complexity of FFT for a vector with length $n$ being $\mathcal O(n\log_2(n))$, the FFT calculation for all the $N_{\mathrm{AR}}N_{\mathrm{loc}}N_{\mathrm{dH}}$ feature vectors in $\mathcal D^t$ is $\mathcal O(N_{\mathrm{AR}}N_{\mathrm{loc}}N_{\mathrm{dH}}N_F\log_2(N_F))$.
Then, the cutoff and normalization for the $N_{\mathrm{AR}}N_{\mathrm{loc}}N_{\mathrm{dH}}$ vectors have the complexity  of $\mathcal O(N_{\mathrm{AR}}N_{\mathrm{loc}}N_{\mathrm{dH}}L_{\mathrm{cut}})$.
Since $L_{\mathrm{cut}}<N_F$, the preprocessing part has computational complexity $\mathcal O(N_{\mathrm{AR}}N_{\mathrm{loc}}N_{\mathrm{dH}}N_F\log_2(N_F))$.

The encoder-decoder neural network part is composed of multiple convolutional and deconvolutional layers.
Based on~\cite{Dumoulin2016AGuide} and~\cite{He2015Convolutional}, the computational complexity of a convolutional/deconvolutional layer given a fixed kernel size can be calculated by $\mathcal O(n_{\mathrm{in}} \cdot n_{\mathrm{out}} \cdot m_{1}\cdot m_{2})$ where $n_{\mathrm{in}}$ and $n_{\mathrm{out}}$ denote the numbers of channels of the input and output data, and $m_1$ and $m_2$ denote the side-lengths of output data.
Thus, based on the network structure information shown in Fig.~\ref{fig: neural net}, the computational complexity of the encoder-decoder neural network part is dominated by those of the Conv~1 and Deconv~1, which are both $\mathcal O(L_{\mathrm{cut}}^2 N_\mathrm{dH}^2 N_{\mathrm{ch}})$.

Finally, in the post-processing part, based on (\ref{equ: anomaly score})$\sim$(\ref{equ: location of anomaly}), the computational complexity is $\mathcal O(N_{\mathrm{ch}})$.
Therefore, in summary, the computational complexity of the sensing algorithm is dominated by those of the preprocessing and encoder-decoder neural network parts, which can be expressed by $\mathcal O(N_{\mathrm{AR}}N_{\mathrm{loc}}N_{\mathrm{dH}}N_F\log_2(N_F)+L_{\mathrm{cut}}^2 N_\mathrm{dH}^2 N_{\mathrm{ch}})$.

\vspace{0.3em}
\textbf{\emph{Remark:}} 
In Section~\ref{ssec: sys comp}, we assume that the {\mms}s are embedded at depth $D_{\mathrm{w}}\geq 0$.
As the received signal model proposed in Section~\ref{ssec: received signal model} and the designed algorithms in Section \ref{sec: joint sensing and transmission optimization} and this section are not dependent on the specific value of $D_\mathrm{w}$, the proposed method is suitable for both $D_{\mathrm{w}}> 0$ and $D_{\mathrm{w}} = 0$ cases, i.e., the embedded and non-embedded scenarios.

\section{Implementation}
\label{sec: system implementation}

In this section, we elaborate on the implementation of the meta-IoT system.
The prototype system is tested in a typical indoor environment and is applied to detecting and locating temperature and humidity anomalies inside walls.
The detailed information of the {\mms}s and the multi-antenna {\rfmeasure} is provided in Sections~\ref{ssec: imp. of mm sensor} and~\ref{ssec: imp. of multi-antenna measure module}, respectively.

\begin{figure}[!t]
\center{\includegraphics[width=0.6\linewidth]{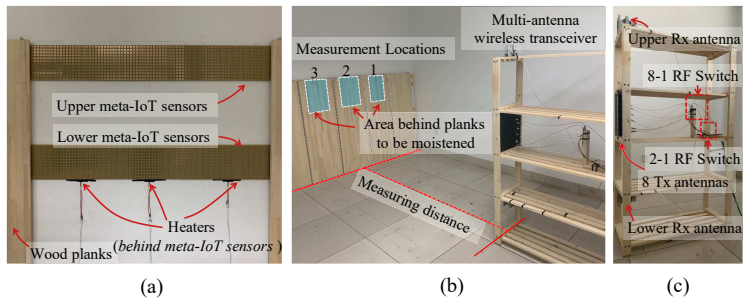}}
	\setlength{\belowcaptionskip}{-1.em}	\setlength{\abovecaptionskip}{-1.em}	\caption{Prototype implementation of the meta-IoT system. (a) Two arrays of designed {\mms}s embedded behind wood planks. (b) The measurement settings deployed in a typical indoor environment, where the {\mms}s are covered by the wood planks, and the multi-antenna {\rfmeasure} is placed at the second measurement location. (c) The multi-antenna {\rfmeasure}, which is installed on a wood frame.}
	\vspace{0.em}
	\label{fig: practical implementation}
\end{figure}

\subsection{{\MmS}s}
\label{ssec: imp. of mm sensor}

Based on the optimal {\mmsd} design obtained in Section~\ref{sec: joint sensing and transmission optimization}, we fabricate the {\mms}s.
Each {\mms} contains $N_\mathrm{T}=2$ {\mmsu}s, and each {\mmsu} has a size of $l_\mathrm{s}\times l_\mathrm{s}\times h_\mathrm{s} = 14.54\times 14.54\times 3.2$ mm$^3$.
The SRRs of the two {\mmsu}s are made of copper lines with $W_{\mathrm{SRR}} = 1.2$ mm width and $H_{\mathrm{SRR}} = 0.035$ mm thickness.
Besides, the side length of the SRRs is $L_{\mathrm{SRR}} = 7.5$~mm, and the gap widths of the two SRRs are $\bm d^* = (1.126, 1.761)$ mm, respectively, which are obtained by solving (\ref{opt: sensor design optimization}).
In a {\mms}, the first {\mmsu} has humidity-sensitive materials filled in its gap, which are the polymer used in the hygristor TELAiRE HS30P~\cite{Hmaterial}.
The second {\mmsu} has temperature-sensitive materials within its gap, which we are the power of thermistor SDNT2012X102-3450-TF~\cite{Tmaterial}.
Moreover, the substrate of each {\mmsu} is FR-4, and the bottom is covered with a thin copper plane to enhance reflection.

As shown in Fig.~\ref{fig: practical implementation}~(a), the {\mms}s are arranged into two rectangle arrays, i.e., an upper {\mms} array and a lower {\mms} array, which are pasted onto a vertical surface.
The upper and lower {\mms} arrays are installed at $h^{\mathrm{MS}}_{1} = 110$ cm and $h^{\mathrm{MS}}_{2} = 58$ cm, respectively, and have a horizontal length of $191.4$ cm and a vertical width of $L_{\mathrm{MS}} = 17.4$ cm.
Each {\mms} array contains $792$ {\mms}s.

To imitate the scenario where the {\mms} arrays are embedded inside the wall, we adopt wood planks to cover the {\mms} arrays.
This setup is based on the experimental observations in~\cite{Ali2003Different}: for signals within $2.4\sim 5$ GHz, wooden walls have a similar attenuation effect with concrete walls.
Besides, according to the standard for engineering acceptance of generic cabling systems~\cite{GBcode}, the cables buried inside a wall should be at least $30$ mm in depth.
Therefore, we use $5$ wood planks with a thickness of $30$ mm to cover the {\mms} arrays, each of which has a size of $120\times 40$ cm$^2$.
The complete setup is shown in Fig.~\ref{fig: practical implementation}~(b), where $N_{\mathrm{loc}} = 3$ measurement locations are considered.

For each measurement location, we consider two types of anomalies, i.e., humidity anomalies and temperature anomalies.
Based on~\cite{Lourenco2017Anomalies}, we imitate the humidity anomalies inside walls by adding a moistened layer in the upper backside region of a wood plank, which is illustrated in Fig.~\ref{fig: practical implementation}~(b).
Specifically, the moistened layer is composed of a piece of wet thin paper, which itself has neglectable influence on the wireless signals.
The vertical length of the moistened layer is $40$ cm, and the horizontal width of the moistened layer is denoted by $W_{\mathrm{M}}=17.5$ cm.

Besides, as in~\cite{Shi2014ANovel}, we control the environmental temperature by using heaters, which are installed behind the lower {\mms} array at each measurement location to generate temperature anomalies.
Given heating power $P_{\mathrm{heat}} = 40$~W, each heater is able to heat the adjacent {\mms}s to be higher than $40^\circ C$ when the wall temperature is $18 ^\circ C$.
Therefore, the temperature difference between the normal and anomalous conditions is larger than $22^\circ C$, which satisfies the standard for testing anomalous conditions of electrical equipments~(i.e., to be larger than $16^\circ C$) as required by National Fire Protection Association~(NFPA)~\cite{NFPAstandard}.

\subsection{Muti-antenna {\RfMeasure}}
\label{ssec: imp. of multi-antenna measure module}

\begin{figure}[!t]
\center{\includegraphics[width=0.5\linewidth]{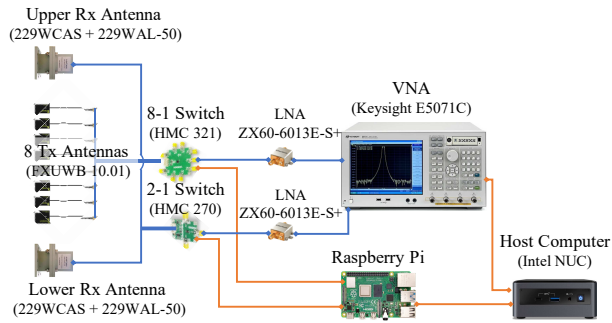}}
	\vspace{-.5em}
	\setlength{\belowcaptionskip}{-1.em}
	\caption{Hardware schematic of the implemented multi-antenna {\rfmeasure}. Blue lines indicate the RF signal links, and the yellow lines indicate the baseband links for data and control signals.}
	\vspace{0.em}
	\label{fig: hardware scheme}
\end{figure}

Fig.~\ref{fig: practical implementation}~(c) shows the implemented multi-antenna {\rfmeasure}.
Besides, its detailed hardware components are shown in Fig.~\ref{fig: hardware scheme} and described as follows.

\subsubsection{Tx Antenna Array}
The Tx antenna array contains $N_{\mathrm{AT}}=8$ Tx antennas, for which we adopt the Taoglas FXUWB10.01 linearly-polarized patch antenna. 
The spacing interval between adjacent Tx antennas is $\delta^{\mathrm{AT}} = 3.67$ cm. 
Besides, the center of the Tx array is at a height of $h^{\mathrm{AT}}_{\mathrm{c}} = 83.6$ cm and is aligned with the center line of the measurement location.
Moreover, a piece of wave-absorbing material is placed behind the Tx antennas.
This setting makes the Tx antennas directional, which satisfies the requirement for the {\rfmeasure} in Section~\ref{sec: overview}.

\subsubsection{Rx Antennas}
We adopt $N_{\mathrm{AR}}=2$ waveguide Rx antennas, each of which is composed of a waveguide-to-coax adapter~(229WCAS) and a rectangular waveguide~(229WAL-50).
The upper and lower Rx antennas are at heights of $h^{\mathrm{AR}}_{1} = 136.3$~cm and $h^{\mathrm{AR}}_{2} = 32$~cm, respectively.

\subsubsection{Radio-frequency~(RF) Switches}
To reduce the number of RF chains needed in the {\rfmeasure}, we adopt two RF switches to multiplex single Tx and Rx chains.
Specifically, the $8$ Tx antennas are connected to a single-pole-$8$-throw RF switch (HMC 321), and the $2$ Rx antennas are connected to a single-pole-$2$-throw RF switch (HMC 270).

\subsubsection{Low Noise Amplifiers~(LNA)}
The poles of the two RF switches are connected to the ports of two LNAs~(ZX-60-43-S+), which are able to provide an average gain of about $13.5$ dB at $[3.5, 4]$ GHz.
The LNAs are able to amplify the transmitted signals sent to the Tx antennas as well as the received signals from the Rx antennas.

\subsubsection{Vector Network Analyzer~(VNA)}
We use a two-port VNA~(Keysight E50171C) as the signal transceiver. 
Specifically, the VAN is set to the forward transmission coefficient, i.e., $\mathrm{S}_{21}$ measurement mode, in order to measure the reflected signals from the {\mms}s.

\subsubsection{Raspberry Pi}
We adopt a Raspberry Pi to control the RF switches selecting which Tx antenna and Rx antenna are connected to the ports of the VNA.

\subsubsection{Host Computer}
The host computer serves as the core controller and data processor of the multi-antenna {\rfmeasure}.
Through Ethernet, it sends control signals to the VNA and the Raspberry Pi and receives the $S_{\mathrm{21}}$ measurement results from the VNA.
Besides, the sensing algorithm in Section~\ref{ssec: unsupervised alg design} is implemented in the host computer by Python.

\section{Evaluation}
\label{sec: evaluation}

\begin{table*}
\centering
	\caption{System Parameters}	
	\label{table: simulation parameters}
	\vspace{-0.8em}
\includegraphics[width=0.91\textwidth]{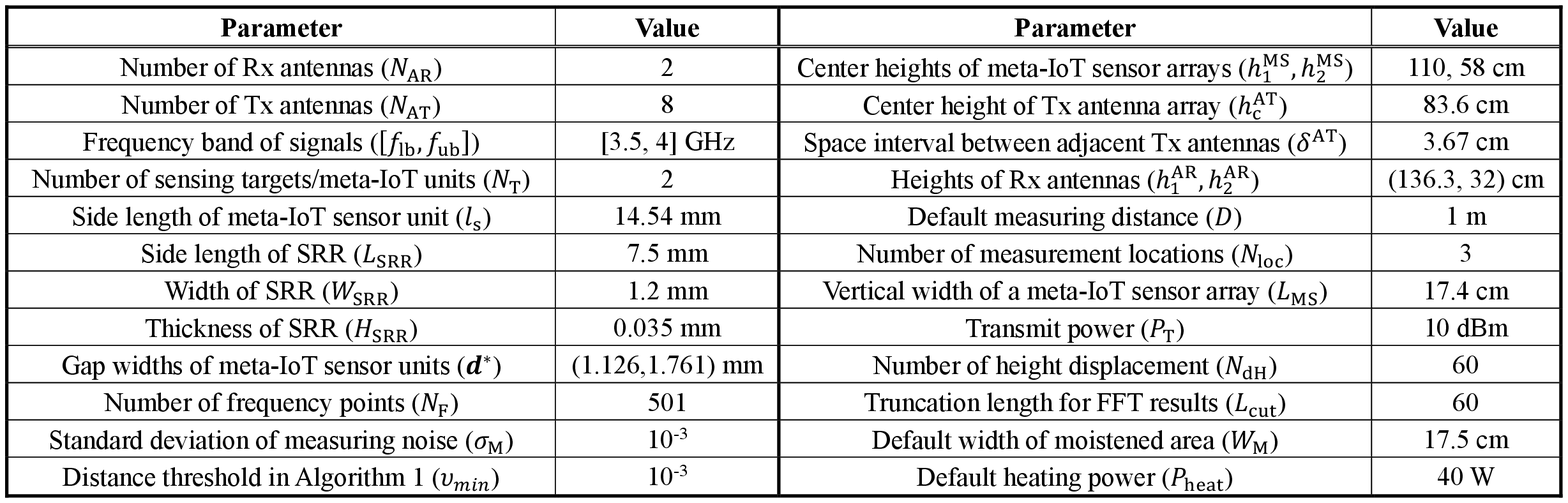}
	\vspace{-.5em}
\end{table*}

In this section, we evaluate the effectiveness of the proposed meta-IoT system.
Specifically, we first provide simulation results for the optimal {\mmsd} obtained by Algorithm~\ref{alg: design optimization algorithm}.
Then, we provide the experimental results for the multi-antenna {\rfmeasure} to validate its capability of sensing {\mms} arrays at different heights.
Following that, we provide the experimental results for the prototype system to detect and locate humidity and temperature anomalies, in the cases of different measuring distances, anomaly scales, noise levels, and depths of the neural network.
The detailed parameters of the system are summarized in Table~\ref{table: simulation parameters}.

\begin{figure}[!t]
\center{\includegraphics[width=0.5\linewidth]{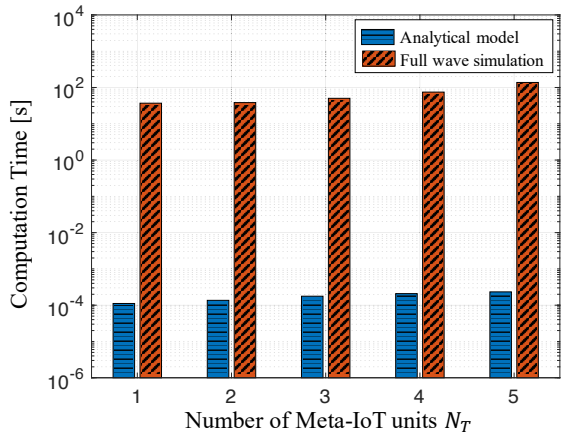}}
    \vspace{-0.5em}
    \setlength{\belowcaptionskip}{-1.em} 
    \caption{Comparison between computation time of reflection coefficient using analytical model and full wave simulation.}
    \vspace{-.5em}
    \label{fig: computation time}
\end{figure}

\begin{figure}[t]
\center{\includegraphics[width=0.5\linewidth]{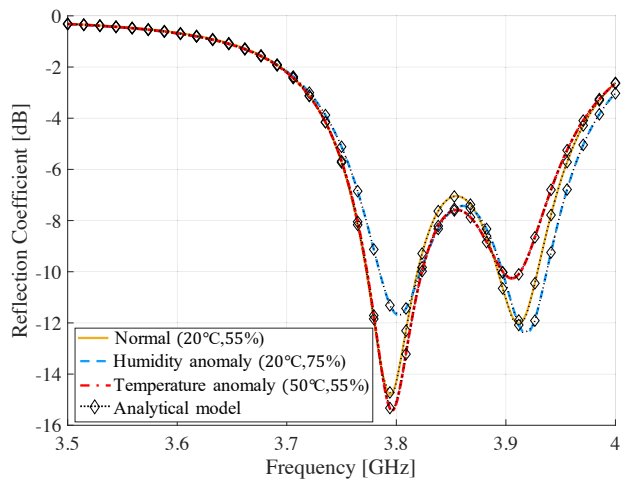}}
	\vspace{-.5em}
	\setlength{\belowcaptionskip}{-0.em}
	\caption{Reflection coefficients of meta-IoT sensor given $\bm d^*$, in the cases of normal, humidity anomaly, and temperature anomaly environmental conditions. The solid and dash lines indicate the results of the full-wave simulation, and the dotted lines with diamond marks indicate the results of the analytical model.}
	\vspace{0.em}
	\label{fig: module 1 performance}
\end{figure}

Firstly, to evaluate the effectiveness of the proposed analytical reflection coefficient model in Section~\ref{ssec: ref coeff model}, we compare its  computation time with that of the full-wave simulation in the CST software.
As shown in Fig.~\ref{fig: computation time}, it can be observed that the computation time of calculating the reflection coefficient using the analytical model is around $5$ orders of magnitude lower than that of the full-wave simulation.

Then, the evaluation of the optimal {\mmsd} obtained by Algorithm~\ref{alg: design optimization algorithm} is provided.
Specifically, when solving Algorithm~\ref{alg: design optimization algorithm}, the same system parameters used in the experimental evaluation in Table~\ref{table: simulation parameters} are adopted.
Besides, the additional parameters in the simulation can be explained as follows. 
The thickness of the wall $D_{\mathrm{w}}=30$ mm, and the relative index of refraction of the wall is $n_{\mathrm{w}}=4.2$ based on~\cite{Ali2003Different}.
Moreover, we consider the performance of meta-IoT sensors in three environmental condition cases, which are the normal case~(temperature $20^\circ$C, humidity $55\%$), temperature anomaly case~(temperature $50^\circ$C, humidity $55\%$), and humidity anomaly case~(temperature $20^\circ$C, humidity $75\%$).
Based on~\cite{Tmaterial,Hmaterial}, the conductivities of the temperature sensitive material in normal and anomaly cases are $0.32$ and $0.97$ [S/m], and those of the humidity sensitive material are $0.11$ and $0.67$ [S/m], respectively.
Furthermore, the available meta-IoT structure set is $\mathcal D_\mathrm{A}=\{ \bm d| d_i\!\in\![0.5,2] $, $d_i\in\mathbb R, i\!=\!1,2 \}$, which ensures the resonant absorption peaks of the two SRRs within $[3.5, 4]$ GHz, and the sampled meta-IoT structure set is $\hat{\mathcal D}_\mathrm{A} = \{ \bm d| d_i\!\in\!\{0.5,1,1.5,2\}, i\!=\!1,2 \}$.
Given the above simulation setting, the optimal {\mmsd} obtained by Algorithm~\ref{alg: design optimization algorithm} is $\bm d^* = (1.126, 1.761)$ mm.

In Fig.~\ref{fig: module 1 performance}, it can be observed that given $\bm d^*$, there exist two distinct reflection coefficient valleys, which are due to the resonance absorption of first and second {\mmu}s, respectively.
Moreover, when a humidity/temperature anomaly happens, the reflection coefficients at the frequencies around the first/second valley become significantly larger, which can be measured by the {\rfmeasure}.
This lays the foundation for the proposed system to detect humidity and temperature anomalies.
Furthermore, the results of the analytical reflection coefficient model in~(\ref{equ: rcs fitting function with polymer}) are in accordance with the results of the full-wave simulation in different cases, which verifies the effectiveness of the analytical model.

\begin{figure}[t]
\center{\includegraphics[width=0.4\linewidth]{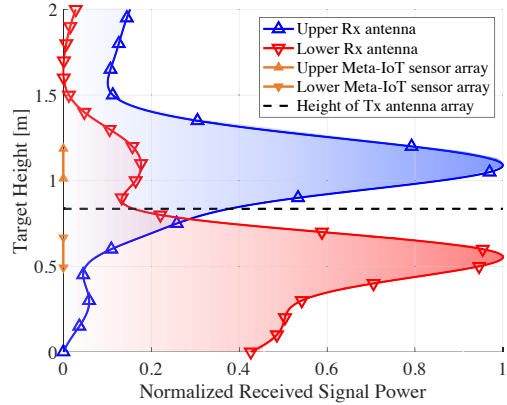}}
	\vspace{-.5em}
	\setlength{\belowcaptionskip}{-1.em}
	\caption{
	Normalized received signal power obtained by the upper and lower Rx antennas when the Tx antenna array applies digital beamforming to steer its beam to different target heights. The received signal power of each antenna is re-scaled through min-max normalization to facilitate observation.
	}
	\vspace{0.em}
	\label{fig: module 2 performance}
\end{figure}

Fig.~\ref{fig: module 2 performance} shows the normalized received signal power obtained by the upper and lower Rx antennas when the digital beamforming is used to extract reflected signals by the {\mms}s at different heights.
It can be observed that when the target height falls into the height ranges of the upper and lower {\mms} arrays, the received signal power reaches its maximum.
This verifies that the designed algorithm can suppress multi-path effects so that the reflected signals by {\mms} arrays at different heights can be obtained separately.
Besides, Fig.~\ref{fig: module 2 performance} also verifies that by setting the added phases according to (\ref{equ: setting varphi}), the arriving phases of the signals which travel from different Tx antennas to the Rx antennas via the meta-IoT sensors at the target height are successfully aligned.
Since the added phases are determined based on the received signal model, Fig.~\ref{fig: module 2 performance} also provides experiential verification for the received signal model in~(\ref{equ: prop channel model equ 1}) proposed in Section~\ref{ssec: received signal model}.

\begin{figure}[!t]
\center{\includegraphics[width=0.6\linewidth]{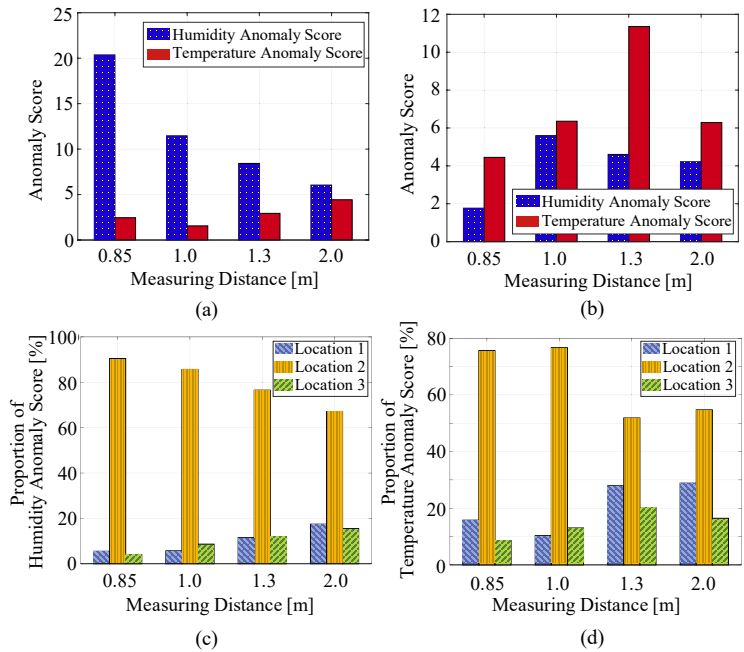}}
	\vspace{-1.5em}
	\setlength{\belowcaptionskip}{-1.em}	\caption{Anomaly scores when (a) a humidity anomaly or (b) a temperature anomaly happens at Location 2, under different measuring distances.
	Besides, (c) and (d) show the proportions of anomaly scores corresponding to different locations under different measuring distances, when the humidity and  temperature anomalies happen, respectively.}
	\vspace{0.em}
	\label{fig: distance influence}
\end{figure}

In Figs.~\ref{fig: distance influence} (a) and (b), it can be observed that within the measuring distance range from $0.85$ m to $2$ m, the accurate anomaly type can be obtained by~(\ref{equ: location of anomaly}), which also indicates that the vertical locations of the anomalies are obtained correctly.
Specifically, when a humidity/temperature anomaly occurs, the humidity/temperature anomaly scores at different measuring distances are larger than the temperature/humidity anomaly scores.
Moreover, from Figs.~\ref{fig: distance influence} (c) and (d), it can be seen that within the range from $0.85$ m to $2$ m, the accurate horizontal location of the happened anomaly, i.e., Location 2, can be obtained correctly.

Furthermore, a larger difference between the anomaly scores indicates a higher capability of the prototype system to detect and locate the anomaly.
In Figs.~\ref{fig: distance influence}~(a) and (c), the difference between the humidity and temperature anomaly scores and the difference between the proportion of humidity anomaly score of Location 2 and those of the other locations become smaller as measuring distance increases.
This indicates that the capability of the prototype in terms of anomaly detection and localization decreases with the measuring distance, which is due to that the received signal power decreases with the measuring distance.
Nevertheless, in Figs.~\ref{fig: distance influence}~(b) and~(d), the capability of the prototype to detect and locate a temperature anomaly is not monotonically decreasing with the measuring distance.
This is because the propagation of the signals reflected by the lower {\mms} array follows a \emph{two-ray} model due to the floor reflection, and thus the received signal power is not in a monotone relationship with the measuring distance. 

\begin{figure}[!t]
\center{\includegraphics[width=0.6\linewidth]{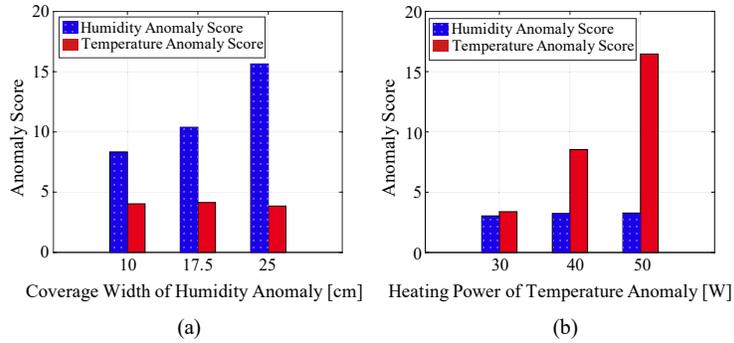}}
	\vspace{-1.9em}
	\setlength{\belowcaptionskip}{-1.em} 
	\caption{Humidity and temperature anomaly scores for (a) a humidity anomaly with different coverage widths, and (b) for a temperature anomaly with different heating power. }
	\vspace{-0.em}
	\label{fig: scale influence}
\end{figure}

Figs.~\ref{fig: scale influence} (a) and (b) show the humidity and temperature anomaly scores for a humidity anomaly with different coverage widths of the moistened layer, and for a temperature anomaly with different heating power, respectively.
It can be seen that given a humidity anomaly or a temperature anomaly with different degrees, the proposed system can judge the type, i.e., the vertical location, of the anomaly correctly.
Besides, with the increment of the coverage width of the humidity anomaly and the heating power of the temperature anomaly, the humidity and temperature anomaly scores increase, respectively.
Therefore, by comparing the relative anomaly scores, the proposed system can potentially estimate the degree of anomaly.

\begin{table*}[!t]
\caption{Comparison of the proposed system with existing works}
\vspace{-1em}
\centering
\tiny 
\label{tab: comparison}
\includegraphics[width=1\textwidth]{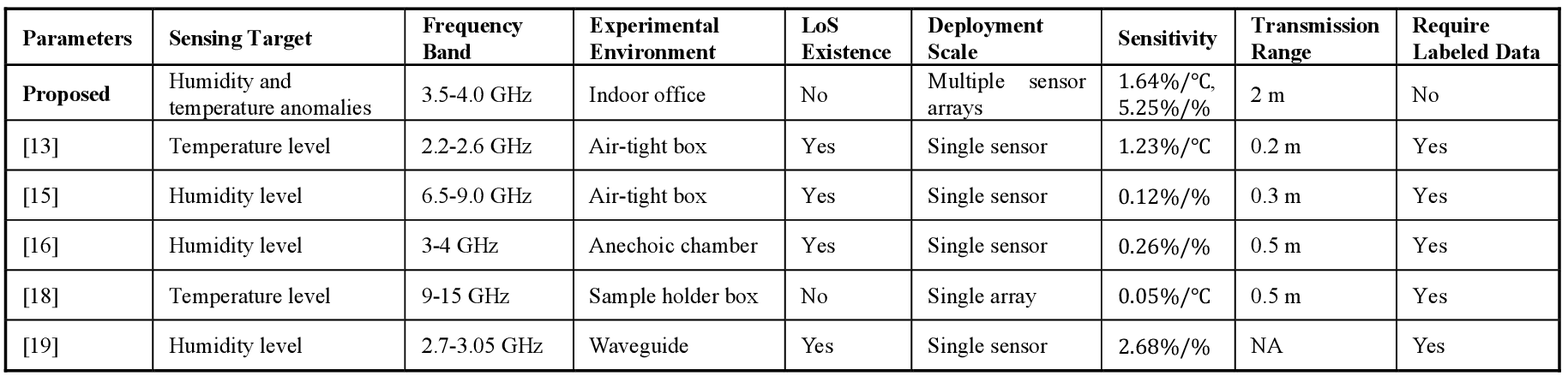}
\vspace{-.5em}
\end{table*}

\begin{figure}[!t]
\center{\includegraphics[width=0.4\linewidth]{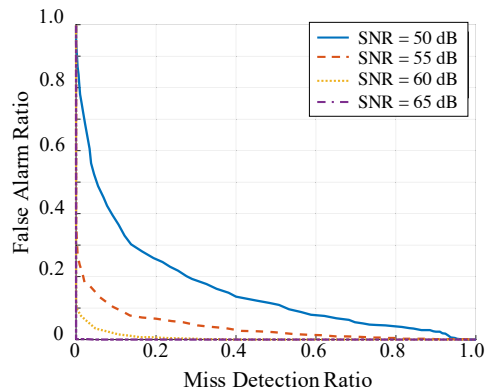}}
	\vspace{-.5em}
	\setlength{\belowcaptionskip}{-1.em} 
	\caption{False alarm ratio versus miss detection ratio of the proposed system under different SNR levels of the received signals.}
	\vspace{0.em}
	\label{fig: noise influence}
\end{figure}

In Fig.~\ref{fig: noise influence}, the false alarm ratio decreases with the miss detection ratio, which can be obtained by increasing the threshold for anomaly detection in~(\ref{equ: occurrence of an anomaly}).
Therefore, there exists a tradeoff in selecting the threshold for anomaly detection.
Moreover, with the increment of SNR of the received signals, both the false alarm ratio and the miss detection ratio can be reduced.
Furthermore, when SNR$\geq 65$ dB, the proposed system is able to detect an anomaly accurately.

\begin{figure}[!t]
\center{\includegraphics[width=0.45\linewidth]{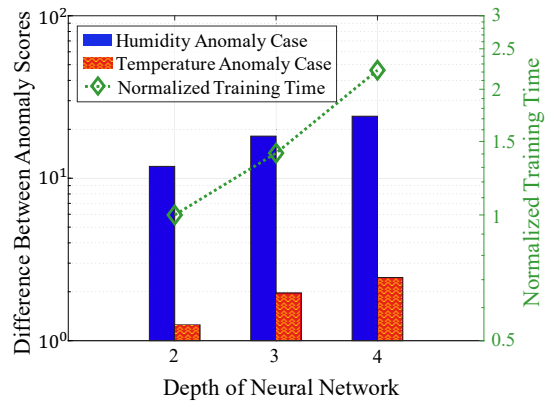}}
	\vspace{-.5 em}
	\setlength{\belowcaptionskip}{-1.em} 	\caption{Difference between anomaly scores and normalized training time versus depth of the encoder-decoder neural network in different anomaly cases, where the minimal training time is normalized to $1$.}
	\vspace{0.em}
	\label{fig: depth influence}
\end{figure}
In Fig.~\ref{fig: depth influence}, as the depth of the encoder-decoder neural network increases, the difference between anomaly scores and the normalized training time have approximately exponential growth.
Nevertheless, the slope of the training time increment is higher than that of the difference between anomaly scores.
Therefore, using an encoder-decoder neural network with a depth larger than $4$ will result in a significant increment of the training time.
This verifies the $4$-leveled encoder-decoder neural network used in the sensing algorithm as it can detect and locate an anomaly accurately without requiring a large amount of training time.

As a summary, we compare the proposed system with the existing works mentioned in Section~\ref{sec: relate work} which use passive chipless sensors working in a similar frequency band and having similar functions.
The comparison results are shown in Table~\ref{tab: comparison}, where the \emph{sensitivity} is defined by the deviation of the minimum reflected coefficient for a unit environmental condition change, i.e., for $1^\circ$C temperature or $1\%$ humidity.
From Table~\ref{tab: comparison}, the following three conclusions can be made.

Firstly, compared with the existing works, the proposed system has higher sensitivities with respect to the humidity and temperature conditions, and achieves a longer transmission range.
This is because the size of a {\mmsu} being smaller than half of the working wavelength enables a larger number of {\mmsu}s to be filled into a given area.
Besides, it makes the reflected signal power of a {\mms} array more concentrated, according to the antenna array theory~\cite{goldsmith2005wireless}.
Secondly, the deployment scale of the proposed system is the largest compared with the existing works, where multiple arrays of sensors are installed to detect and locate anomalies within a large region.
Therefore, the proposed system is more suitable for 6G, which requires the sensors to be densely deployed on a large scale.
Thirdly, the proposed system does not rely on labeled data, since its sensing algorithm is trained in an unsupervised manner.
Thus, the proposed system especially fits for the applications of anomaly detection and localization, since the anomalous data are generally unavailable in advance.
 
\section{Conclusion}
\label{sec: conclu}
In this paper, we have considered a meta-IoT system with a large number of {\mms}s, and have proposed a joint sensing and transmission design for this system.
Specifically, we have first modeled the signal at the receiver considering the joint influence of sensing and transmission.
Based on this model, we have optimized the meta-IoT sensor design for the sensing performance.
Then, we have designed an unsupervised sensing algorithm based on a convolutional encoder-decoder neural network, which can robustly sense environmental anomalies.

Based on the experimental evaluations of the prototype, we can offer the following conclusions.
\begin{itemize}[leftmargin=*]
\item Firstly, compared with other existing works, the designed system has a higher sensitivity and a longer transmission range.
The prototype system has demonstrated the capability to successfully detect and locate anomalies within $2$ m and distinguish the locations of anomalies with a spatial resolution of $0.4$ m.
\item Secondly, by using the designed sensing algorithm, multi-path effects can be suppressed so that the signals reflected by different {\mms} arrays can be obtained separately.
\item Thirdly, the proposed system has been shown to be robust to noises.
To be specific, the false alarm and miss detection ratios of the prototype system decrease with the SNR, and approach zero for SNRs larger than $65$~dB, given the typical indoor environment.
\end{itemize}

 \begin{appendices}
 \section{Proof of Proposition~\ref{prop: channel model}}
\label{appx: derivation of large number channel model}

Since a larger number of {\mms}s can be collectively considered as a uniform effective medium, the {\mms} array performs specular reflection according to the principles of geometric optics.
The equivalent reflection coefficient is denoted by $\chi(f, \bm d, \bm c, D)$.
Thus, (\ref{equ: prop channel model equ 1}) can be derived.
Moreover, $\chi (f, \bm d, \bm c, D)$ can be calculated by the following special case.
For an infinite {\mms} array surface, consider that a Tx antenna and a Rx antenna which are superimposed and both at distance $D$ from the {\mms} array.
Then, based on general channel model (\ref{equ: int channel model}), the received signal at the Rx antenna can be calculated by
\begin{small}
\begin{align}
& y(f, \bm d, \bm c, D) = \sqrt{P}\cdot g^{\tx}(f) \cdot g^{\rx}(f) \cdot \gamma(f, \bm d, \bm c)  \cdot \iint_{-\infty}^{\infty} \big( \frac{v e^{-\sqrt{D^2+y^2+z^2}(\beta \frac{D_\w}{D} + \iu(\frac{2\pi f}{v} + \frac{2\pi f D_\w}{vD}(n_\w-1)) )}}{4\pi f\sqrt{D^2 + y^2 + z^2}}\big)^2 \d y\d z \nonumber \\
\label{equ: appendix A equ}
& = \sqrt{P}\cdot g^{\tx}(f) \cdot g^{\rx}(f) \cdot \gamma(f, \bm d, \bm c) \cdot \frac{v^2}{16\pi^2 f^2} \cdot 2\pi \cdot \mathrm{Ei}\big(2D\cdot ( \beta \frac{D_\w}{D} + \iu(\frac{2\pi f}{v} + \frac{2\pi f D_\w}{vD}(n_\w-1))  ) \big) 
\end{align}
\end{small}
where $\mathrm{Ei}(\vartheta)$ denotes the exponential integration function.
Based on the property $\mathrm{Ei}(\vartheta) = \frac{e^{-\vartheta}}{\vartheta} \sum_{k=0}^{K-1} \frac{k!}{(-\vartheta)^k}$, when $D$ is large, (\ref{equ: appendix A equ}) can be approximated by
\begin{small}
\begin{align}
 y(f, \bm d, \bm c, D) &\approx \sqrt{P}\cdot g^{\tx}(f) \cdot g^{\rx}(f) \cdot \underbrace{\frac{D v^2\gamma(f, \bm d, \bm c)}{2\beta {D_\w} v f \!+\! 4\pi f^2\iu (D \!+\! {D_\w} (n_\w \! -\! 1 ))}}_{\text{Coefficient } \chi(f, \bm d, \bm c, D)}  \cdot \underbrace{\frac{ve^{-(\beta \frac{D_\w}{D} \!+ \iu(\frac{2\pi f}{v} \!+\! \frac{2\pi f D_\w}{vD}(n_\w-1))  )\cdot 2D}}{4\pi f\cdot 2D}}_{\text{Specular Reflection } g_{\text{sr}}(\bm x^\rx, \bm x^\tx, f, D)},\nonumber
\end{align}
\end{small}
which proves (\ref{equ: prop A chi}) in Proposition~\ref{prop: channel model}.$\hfill\blacksquare$\par

 \end{appendices}

\begin{small}
\bibliographystyle{IEEEtran}
\bibliography{ms}
\end{small}

\end{document}